\newcommand{\secref}[1]{Sec.~\ref{#1}}
\newcommand{\figref}[1]{Fig.~\ref{#1}}

\newcommand{\equref}[1]{Eq.~(\ref{#1})}

\documentclass[lettersize,journal]{IEEEtran}
\usepackage{amssymb}
\usepackage{amsmath}
\usepackage{graphicx}
\usepackage{multirow}
\usepackage{algorithm}
\usepackage{algpseudocode}
\usepackage{epstopdf}
\usepackage{amsmath}
\usepackage{algorithmicx}
\usepackage{algpseudocode}
\usepackage{array}
\usepackage{color}
\usepackage{soul}
\usepackage{xcolor}
\usepackage{graphics}
\usepackage{multirow}
\usepackage{setspace}   
\usepackage{enumitem}
\usepackage{epstopdf}
\usepackage{pdfpages}
\usepackage{amssymb}
\usepackage{graphicx}
\usepackage{makecell}
\usepackage{enumitem}
\usepackage{booktabs}
\makeatletter

\newcommand{\Rmnum}[1]{\expandafter\@slowromancap\romannumeral #1@}

\newtheorem{remark}{Remark}

\usepackage{hyperref}
\hypersetup{
	colorlinks=true,   
	linkcolor=magenta,     
	citecolor=teal,   
	filecolor=mycustompurple,   
	urlcolor=red    
}

\makeatother

\usepackage{balance}

\begin{document}
	
	\title{From Reliability to Security: How RIS-Assisted Adaptive SM and SSK Enhances Wireless Systems}
	
	\author{Chaorong Zhang,
		Benjamin K. Ng,
		Ke Wang,
		Hui Xu, and\
		Chan-Tong Lam}
	
	\maketitle
	
	\setlength{\parskip}{-1pt}

	\begin{abstract}
		This paper proposes two novel wireless transmission schemes, namely reconfigurable intelligent surface (RIS)-assisted received adaptive spatial modulation (RASM) scheme and RIS-assisted received adaptive space shift keying (RASSK) scheme, designed to enhance spectral efficiency (SE) and physical layer security (PLS).	
		In both proposed schemes, transmitting bits are dynamically mapped at receive antennas by leveraging the characteristics of the RIS in each time slot,
		which enables the enhancement of signal-to-noise ratio (SNR) at specific selected antennas \textcolor{magenta}{with nearly few power, thus leading to a reliable} and green wireless communication. 
		\textcolor{magenta}{Unlike conventional fixed-antenna RIS-RSM/GSSK, the term ``adaptive'' indicates the number of active antennas dynamically changes per symbol, conveying extra spatial information to break existing spectral efficiency bottlenecks.}
		This adaptive approach facilitates the conveyance of extra bits to the receiver, which means it needs less cost of radio-frequency chains at transmitter while improving SE.
		Besides, the proposed schemes offer an inherent PLS security advantage, as the eavesdropper is unable to completely detect signals reflected from the RIS.
		To comprehensively evaluate the performance of the proposed RASM and RASSK schemes, this paper presents a detailed analytical performance of their spectral efficiency, detection complexity, bit error rate, and secrecy rate,  which are accompanied by insightful findings and conclusions.
		Simulation and analytical results demonstrate the superiority of the proposed schemes, showcasing their improved error performance and robustness against wiretapping, while also highlighting the potential of the RASM and RASSK schemes for future wireless applications. 
	\end{abstract}

	
	\IEEEpeerreviewmaketitle
	
	\vspace{-6pt}
	
	\section{Introduction} \label{sec 1}
	
	\subsection{Index Modulation} 
	With the exponential growth of devices accessing wireless networks, communication throughput and data rates have surged recently. 
	In response, novel techniques have emerged, with index modulation (IM) standing out. 
	Numerous variants exist in the IM area. Prominent multi-antenna techniques include spatial modulation (SM), space shift keying (SSK), generalized space shift keying (GSSK), and others.
	These traditional schemes map antenna indices to convey information: single-antenna activation in SM/SSK and multi-antenna combinations in generalized schemes, enabling extra bits and enhancing spectral efficiency (SE) \cite{ref1}. 
	In IM schemes, constellation symbols modulated by $M$-ary phase shift keying (PSK) or quadrature amplitude modulation (QAM) are transmitted, while additional bits are encoded via antenna indices, each representing distinct transmitted bits. 
	Unlike SM, SSK omits $M$-ary symbols, reducing digital modulation complexity at transmitter and receiver. As noted in \cite{ref2}, SSK exhibits slightly better bit-error-rate (BER) performance than SM with perfect channel state information (CSI) at the receiver. However, conventional SSK \cite{ref2} demands numerous antennas for higher SE, raising costs and degrading BER. 
	GSSK, introduced in \cite{ref3}, alleviates antenna costs by activating a fixed number, but yields inferior BER compared to SM and SSK. 
	Thus, another groundbreaking IM approach is called \textcolor{magenta}{adaptive} spatial modulation (ASM), which provides flexibility through variable activated transmit antennas forming diverse antenna combinations (ACs), achieving superior SE over traditional SM, SSK, etc. \cite{ref4}. 
	By minimizing transmit antennas or ACs, ASM's BER outperforms other IM schemes at equivalent SE. 
	ASM incorporates constellation symbols, increasing complexity relative to adaptive space shift keying (ASSK), which excludes them. 
	To lower the processing complexity at transmitter, this paper also explores ASSK, based on SSK principles where antennas emit pure power, simplifying modulation. 
	As a trade-off, ASM and ASSK enhance SE beyond traditional IM schemes but introduce challenges, such as elevated radio frequency (RF) chain costs and physical layer security (PLS) concerns. To address these, a novel technique is applied to ASM and ASSK for system improvement.
	
	\subsection{Reconfigurable Intelligent Surface}
	As a promising technique garnering considerable attention in recent years, reconfigurable intelligent surfaces (RIS) holds great potential for application in future beyond 5th generation (B5G) and 6th generation (6G) wireless communication systems. 
	The RIS comprises a chip, referred to as the intelligent unit, which is connected to the transmitter.
	This intelligent unit receives information about phase adjustments and subsequently controls the angles of the reflection elements (REs) accordingly \cite{ref5}. 
	The REs, as fabricated from the special materials that can reflect electromagnetic waves \cite{ref6},  are arranged in a two-dimensional (2D) planar array \cite{ref7} and can be individually controlled by the intelligent unit. 
	According to it, a deployment feature in the RIS is provided, that is, an extension in coverage of wireless communication through altering the communication delivery into the desired directions \cite{ref8}. 
	\textcolor{red}{Recently, the state-of-the-art applications of RIS have expanded profoundly, integrating advanced architectures, such as multi-layer refracting and self-powered absorptive surfaces, into complex space-air-ground integrated networks and satellite-terrestrial systems to fundamentally boost both coverage reliability and endogenous physical layer security \cite{RR_R1_A, RR_R1_B, RR_R1_C, RR_R1_D}.
	Unlike the traditional relays, the RF chains and amplifiers are not required in the RIS, which can better improve the power utilization in communication. }
	Therefore, through the RIS, the phase shift of transmitted signals can be adjusted with near few powers and the signal-to-noise ratio (SNR) can be further enlarged at receiver, which is named the passive beamforming \cite{ref9}. 
	By possessing these advantages, the RIS can achieve the IM scheme at receive antennas by adjusting the phase to achieve the passive beamforming at one or more specific antennas, 
	in which it can more effectively address the aforementioned issues of the ASM and ASSK schemes mentioned above and provides a viable consideration of system design for the proposed scheme, as illustrated as follows.

	\vspace{-10pt}

	\subsection{RIS-assisted IM schemes}
	Given extensive research on reconfigurable intelligent surfaces (RIS), IM is increasingly integrated into its applications. 
	First, RIS acts as a passive relay to enhance IM communication quality. In \cite{ref10}-\cite{ref11}, an RIS-assisted IM scheme is proposed where SSK operates at transmit antennas, improving BER performance via RIS assistance, as termed TSSK herein for convenience. 
	RIS-assisted SM schemes are explored in various state-of-the-art works \cite{ref12}-\cite{ref16}, with SM at transmit antennas and RIS boosting receiver SNR, \textcolor{magenta}{so-called the TSM in this paper.} 
	To reduce antennas, RIS-assisted GSSK is introduced in \cite{ref12}, adhering to GSSK activation principles at transmit antennas, though with inferior BER compared to TSSK and TSM. 
	Recent advancements include diverse RIS-assisted IM variants operable not only at transmitters but also at receivers via antenna selection, referred to as received IM herein. 
	As a trade-off in RIS-assisted systems, received IM offers advantages like superior BER, reduced transmitter complexity, and potential gains in SE over related schemes \cite{ref17}-\cite{ref21}. 
	In \cite{ref17}, RIS-assisted RSM and RSSK are first proposed with antenna selection at receive antennas, yielding good BER but omitting transmitter-RIS channel due to AP design, with room for SE improvement. 
	Subsequently, RIS-assisted RGSM and RGSSK schemes emerge in \cite{ref18}-\cite{ref19}, enhancing SE via novel low-complexity detectors, albeit with poorer BER than RSM and RSSK. 
	All these schemes employ fixed selected or activated antennas for information mapping, limiting flexible SE enhancement or dynamic antenna adjustment. 
	To overcome this, we propose two novel schemes with a flexible adaptive antenna selection algorithm, enabling dynamic resource configuration for improved SE adaptability and surpassing fixed-antenna constraints. 
	\textcolor{magenta}{While our proposed schemes rely on coherent detection, acquiring the requisite cascaded CSI typically requires dedicated pilot-based training protocols. To circumvent this estimation overhead, non-coherent transmission designs present a promising alternative \cite{RR_R3_A, RR_R3_B, RR_R3_C}, which demonstrate significant potential in reducing system complexity while enhancing secure communications.}
	\textcolor{green}{Unlike classic SM/SSK that requires multiple power-hungry transmit RF chains and relies on random fading, receive-side RIS-IM uses a single-antenna transmitter and a RIS to artificially reshape the channel. This paradigm minimizes IoT transmitter costs and enhances security. Advancing this, our schemes introduce an adaptive receive-antenna activation mechanism, fundamentally breaking the SE bottlenecks of conventional fixed-antenna RIS-IM.}
	\textcolor{red}{Besides, differing from traditional received IM schemes that rely on a fixed number of active antennas and physically deactivate the rest, our proposed schemes keep all receive antennas active and utilize RIS passive beamforming to dynamically vary the number of selected antennas per time slot. 
	This adaptive mechanism fundamentally breaks the SE bottlenecks of fixed-antenna designs while intrinsically avoiding the computational overhead of complex real-time channel optimization. }
	The RIS-assisted transmitted ASM scheme is termed TASM, with ASM at the transmitter.

	\vspace{-10pt}

	\subsection{Contributions}
	The existing RIS-assisted IM schemes still face several critical challenges, e.g.,  contradicting with the low-cost nature of wireless communications to multiple RF chains, limit SE adaptability in dynamic channels by fixing antenna activation patterns, low PLS in wireless applications, etc. 
	To address these issues, we propose two novel RIS-assisted received IM schemes, \textcolor{magenta}{termed the RIS-assisted received ASM} (RASM) and RIS-assisted RASSK schemes, which dynamically map information bits to receive ACs instead of transmit antennas. 
	Note that prior works on RIS-assisted IM schemes primarily focus on BER or data rate, overlooking the potential of RIS to enhance PLS in this domain.
	By leveraging RIS’s passive beamforming to enhance SNR at selected receive ACs, our schemes achieve higher SE, while the unique design in reflections inherently enhances the physical layer security against eavesdropping.
	Overall, the main contributions of this work can be briefly summarized as follows:
	\begin{itemize}
		\item This paper proposes two novel RIS-assisted IM schemes, denoted as RASM and RASSK, which enhance performance metrics such as BER, spectral efficiency, secrecy rate (SR), and robustness by first employing adaptive antenna selection at the receive antennas and leveraging the RIS to boost the SNR at the receiver.
		\item In our schemes, eavesdroppers face challenges in reconstructing transmitted signal due to the inherent security mechanism embedded in the dynamic receive-side mapping. This paper pioneers the theoretical analysis of the SR for these schemes under a passive eavesdropper, bridging the gap in PLS-based studies for RIS-assisted IM schemes.
		\item Accounting for side lobes and inevitable interference, the REs are partitioned into constructive and non-constructive components in our schemes with detailed analytical performance. Also, to further reduce the complexity of the proposed schemes, a generalized pre-defined ACs selection at the receiver is designed.
		\item This work presents Monte Carlo simulations of the proposed schemes, evaluating BER, SE, and secrecy rate under perfect and imperfect CSI scenarios with artificial noise (AN), ML and maximal ratio combining (MRC) detectors, yielding valuable insights.
	\end{itemize}
	\vspace{-10pt}
	\subsection{Organization and Notation}
	\textit{Organization:} The rest of the paper is organized as follows: 
	In Sec. \Rmnum{2}, the system model of the RASM and RASSK schemes with introduction of detection is given. 
	Sec. \Rmnum{3} provides the illustration of random ACs selection.
	The analytical performance of SE, BER performance, and detection complexity is presented in Sec. \Rmnum{4}. 
	Sec. \Rmnum{5} provides the system model of the RASM and RASSK schemes with the eavesdropper and the analytical performance in PLS of the proposed schemes by deriving SR. 
	The theoretical and simulation results of the RASM and RASSK schemes and other related schemes with different conditions are shown in Sec. \Rmnum{6}. 
	Eventually, the conclusion is given in Sec. \Rmnum{7}. 
	\par
	\textit{Notation:} Boldface letter is used to denote the matrices. 
	$\left( \cdot \right) ^\mathsf{T}$ and $\left( \cdot \right) ^\mathsf{H}$ denote the transpose and complex conjugate transpose operations. 
	$\mathsf{diag}(\cdot )$ represents diagonal matrix operation. $\mathbb{C} ^{n\times m}$ is the space of $n\times m$ complex-valued matrices. 
	The real part of a complex variable can be denoted as $\mathfrak{R} \{\cdot \}$. 
	$\mathrm{P_r}\left( \cdot \right)$ represents the probability of an event. 
	The exponential function is denoted by both $\exp \!\:(\cdot )$ and $e^{\left( \cdot \right)}$, which, while distinct in notation, convey the identical mathematical concept.
	$\mathrm{Q(}\cdot )$ means the Q-function.
	$\left\| \cdot \right\| _2$ is the second-norm operation, while $\mathbb{E} (\cdot )$ stands for statistical expectation operation. $\jmath\triangleq\sqrt{-1}$ is the imaginary number. 
	$f\left( \cdot \right) $ and $f\left( \cdot \middle| \cdot \right) $ denote the probability density function (PDF) the conditional PDF. 
	$\mathbb{E} \left( \cdot \right) $ represents the operation of expectation. 
	$\left( \begin{array}{c}	n\\	k\\\end{array} \right) $ stands for the binomial coefficient from $n$ choose $k$. $\mathcal{C} \mathcal{N} (\mu ,\sigma ^2)$ represents the complex Gaussian distribution with mean $\mu $ and variance $\sigma ^2$.
	$\mathbb{D}(\cdot )$ stand for variance operations.
	
	\begin{figure*}
		\centering
		\includegraphics[width=16.5cm,height=7.5cm]{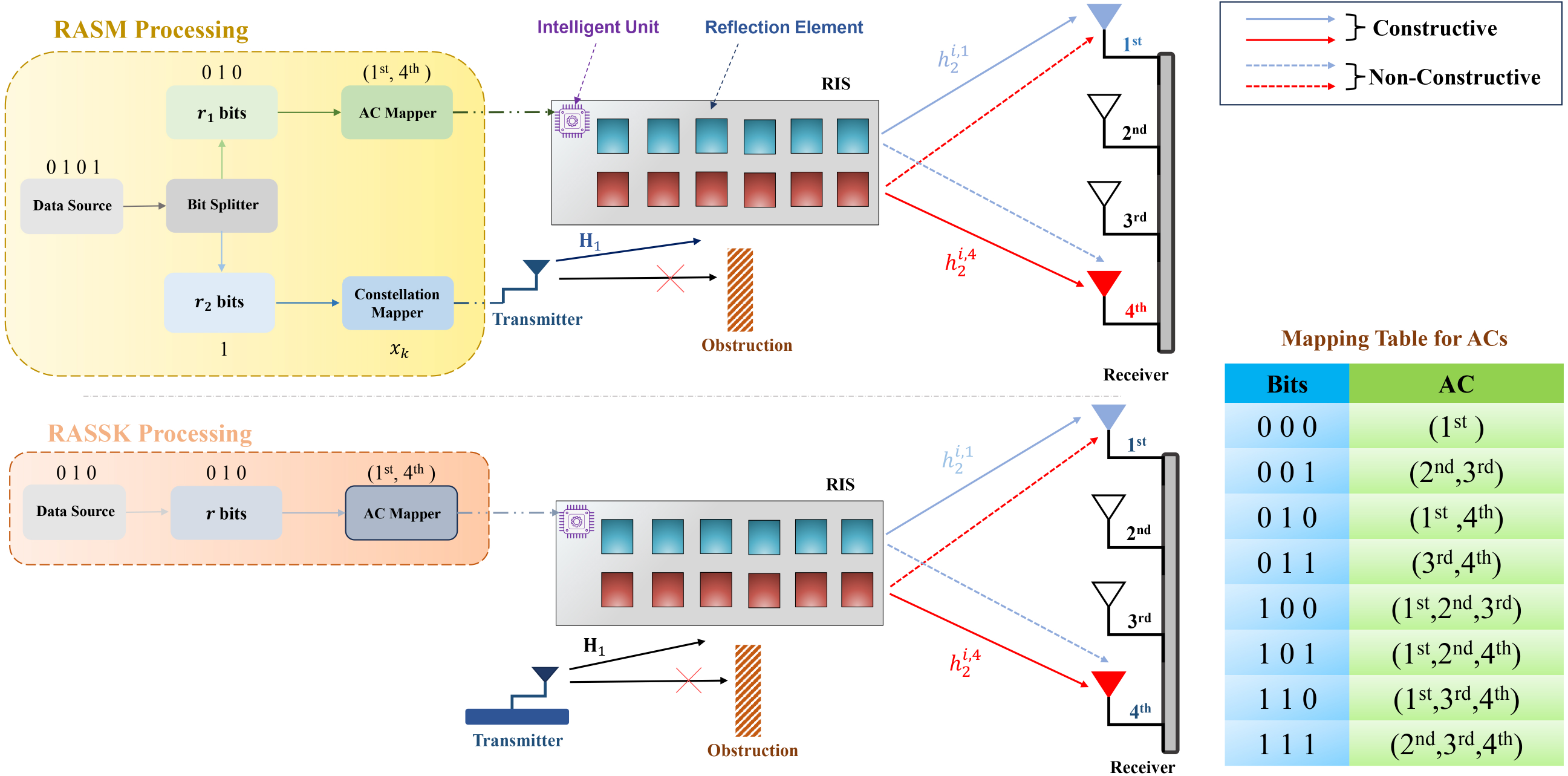}\\
		\caption{System models for the RASM and RASSK schemes.}
		\label{system model}
	\end{figure*}
	
	\section{System Model}  \label{sec 2}
	\subsection{Configuration and Wireless Channel}

	\subsubsection{Applied Scenarios}
	This section illustrates the system models for the proposed RIS-assisted received IM schemes, RASM and RASSK, to enhance wireless networks. As depicted in \figref{system model}, we consider a SIMO uplink in an IoT scenario \textcolor{red}{where fixed sensors collect environmental data} \footnote{\textcolor{red}{Since both transmit sensors and the receive BS are fixed in this scenario, mobility-induced channel fluctuations are beyond the scope of this paper.}}, e.g., temperature, and transmit it to a base station (BS). 
	The sensor employs a single transmit antenna ($N_t=1$), while the BS has $N_r$ receive antennas. 
	Due to obstructions, the direct link is blocked, relying solely on the RIS path. 
	The RIS comprises $N$ REs and an intelligent unit linked to the transmitter.
	For instance, in the RASM scheme, bit sequence 010 maps to an AC selecting the first and fourth receive antennas; the intelligent unit directs the first RE subset to amplify the signal toward the first antenna and the second subset toward the fourth. Note that RASM transmits $M$-ary constellation symbols, whereas RASSK conveys only pure power, necessitating more receive antennas in RASSK. We denote $N_r^1$ for RASM and $N_r^2$ for RASSK, with $N_r^1 \ll N_r^2$.

	\subsubsection{Rayleigh Channel Fading}
	In both system models, the channel matrix of the channel from the transmitter to the receiver is defined as $\mathbf{H}_1\in \mathbb{C} ^{N\times 1}$, 
	the other from the RIS to the receiver is represented as $\mathbf{H}_2\in \mathbb{C} ^{N_r\times N}$, which both follow the Rayleigh flat fading with zero mean value and unit variance. 
	The diagonal matrix of adjusted phase in the RIS can be given as $\mathbf{\Phi }=\mathsf{diag}(\exp \!\:\left( \jmath \varphi _{1,m} \right) ,\exp \!\:\left( \jmath \varphi _{2,m} \right) ,\cdots ,\exp \!\:\left( \jmath \varphi _{N,m} \right) )$, with $\varphi _{i,m}$ representing the adjusted phase shift of $i$-th reflecting element and for $m$-th receive antenna. 
	Furthermore, we define the wireless channel from the transmit antenna to the $i$-th reflecting element$\mathrm{ }$ and the one from the $i$-th reflected to the $m$-th receive antenna as $h_{1}^{i}$ and $h_{2}^{i,m}$ with $i=1,2,\cdots ,\mathrm{ }N$ and $m=1,2,\cdots ,\mathrm{ }N_r$, respectively. 
	$h_{1}^{i}$ and $h_{2}^{i,m}$ are assumed to be independent and identically distributed complex Gaussian random variables with $\mathcal{C} \mathcal{N} (0,1)$, which can be extended to $h_{1}^{i}=\alpha _ie^{-\jmath\theta _i}$ and $h_{2}^{i,m}=\beta _{i,m}e^{-\jmath\omega _{i,m}}$, with $\alpha $ and $\beta $ representing the Rayleigh factors in various channels, 
	as well $\theta $ and $\omega $ representing the phase shift of these three channels. 

	\subsubsection{Specific Antenna Settings}
	\textcolor{magenta}{For convenience, we define the receive antenna activation vector for the $r$-th AC as $\boldsymbol{v}_r = [v_{r,1}, v_{r,2}, \dots, v_{r,N_r}]^{\mathsf{T}} \in \{0,1\}^{N_r \times 1}$, where $v_{r,n} = 1$ indicates that the $n$-th receive antenna is selected, and $v_{r,n} = 0$ otherwise. 
	For example, based on the mapping table in \figref{system model}, if the transmitted bit sequence is `001', the second and third antennas are activated. The corresponding receive antenna vector is thus denoted as $\boldsymbol{v}_2 = [0, 1, 1, 0]^{\mathsf{T}}$. 
	The number of active antennas in each AC, denoted as $N_a$ ($1 \le N_a \le N_r$), dynamically changes within each symbol period. 
	For instance, as shown in \figref{system model}, $N_a = 1$ is employed to transmit the bit sequence `000', whereas $N_a = 2$ is utilized for the sequences `001' and `010'. 
	This dynamic activation mechanism significantly enhances the flexibility of antenna configurations and substantially improves the SE. 
	The full set of permissible ACs is defined as $\varLambda$, containing $J$ possible combinations. 
	Consequently, the total number of ACs can be expressed as $J = \sum_{n=1}^{N_r} \binom{N_r}{n} = 2^{N_r} - 1$. 
	Following the principle of the proposed adaptive scheme, although $N_r$ receive antennas are deployed, a dynamically varying subset of $N_a$ antennas is selected to convey the spatial information. }
	
	\subsubsection{Working Principle}
	The REs are evenly divided into $N_a$ parts, which has $N_E=\lfloor \frac{N}{N_a} \rfloor $ number of REs, and each part is corresponding for one selected receive antenna to achieve the passive beamforming. 
	\footnote{This even division strategy for RE allocation ensures fair resource distribution without real-time optimization. It enables straightforward passive beamforming by assigning equal RE subsets to boost SNR at each selected antenna. The method reduces computational overhead at the intelligent unit, suiting dynamic IoT scenarios with limited processing power.}
	Besides, each antenna not only can receive the signals reflected from the REs that are corresponding for itself but also from the non-corresponding ones. 
	Specifically, as shown in Fig.1, the first receive antenna can receive the signals reflected from the REs corresponding for itself (as shown in the red dashed line) and can be affected by those for the fourth receive antenna (as shown in blue dashed line) in the meanwhile, 
	which leads to the constructive part and non-constructive part in the signals for $m$-th receive antenna, respectively.  
	Notably, all $N_r$ receive antennas remain active during transmission, unlike traditional activation-based IM where unselected antennas are deactivated. 
	Instead, the RIS employs passive beamforming to focus gain through constructive reflections on selected antennas while imposing non-constructive interference on others. 
	The RE allocation strategy involves sequential assignment based on the sorted order of selected receive antenna indices in the AC, pre-agreed upon by transmitter and receiver. 
	As shown in the mapping table in \figref{system model}, for bit sequence 010 (mapping to antennas 1 and 4), the first RE part amplifies the signal toward antenna 1, and the second toward antenna 4. 
	Similarly, for 100 (mapping to antennas 1, 2, and 3), the first part targets antenna 1, the second antenna 2, and the third antenna 3. 
	This sequential approach enables straightforward implementation at intelligent unit, which parses the bit sequence to identify the AC and sorts indices before phase adjustment. 
	\textcolor{red}{It ensures fairness by allocating equal RE resources to each selected AC, strictly avoiding the overhead associated with real-time dynamic channel optimization algorithms, ensuring high efficiency.}

	\subsubsection{Rician Channel Fading}
	To better model a practical wireless environment for the proposed schemes, the Rician channel fading can also be taken into account. 
	This type of fading is appropriate when there is a strong line-of-sight (LoS) path between the RIS and the receiver.
	The Rician distribution of $h_{2}^{i,n}$ can be given as
	\begin{align}
		h_{2}^{i,n}=\sqrt{\frac{K}{K+1}}h_{2,Los}^{i,n}+\sqrt{\frac{1}{K+1}}h_{2,NLos}^{i,n},
	\end{align}
	where $h_{2,Los}^{i,n}$ represents the deterministic LoS component and $h_{2,NLos}^{i,n}$ is a zero-mean complex Gaussian random variable accounting for the non-LoS multi-path components. 
	The Rician factor $K\geqslant 0$ denotes the ratio of the power in the LoS component to that in the scattered components. 
	When $K= 0$, the model reduces to Rayleigh fading, while higher $K$ values indicate a stronger LoS presence.
	
	\vspace{-10pt}
	\subsection{RASM Transmission} 
	Assuming total $b_{\mathrm{RASM}}$ bits are required to transmit, $b_{\mathrm{RASM}}$ bits can be divided into $b_1$ and $b_2$ bits in the RASM scheme, 
	where $b_1$ bits are mapped into the indices of selected antenna combinations at receiver and $b_2$ bits are mapped into the ones mapping to $M$-ary PSK/QAM symbol transmitted from the transmitter, as shown in \figref{system model}.
	For the selected AC, each one represents different bit sequence, e.g., the third selected AC indicates the bits 010.
	Moreover, the last bit in sequence is 1, which is represented by $k$-th constellation symbol $x_k$ in $M$-ary PSK/QAM constellation symbol vector as $\boldsymbol{x}_k=\left[ x_1,x_2,\cdots ,x_M \right] ^\mathsf{T}$ as shown in \figref{system model}. The expression for the received signal in the RASM scheme can be written as
	\begin{align} \label{RASM recevied signal}
		\boldsymbol{Y}=\mathbf{H}_2\mathbf{\Phi }_r\mathbf{H}_1x_k+\mathbf{N},
	\end{align}
	where $\mathbf{\Phi }_r \in \mathbb{C} ^{N\times N}$ represents the diagonal matrix of adjusted phase for $r$-th AC is the vector of Gaussian white noise. 
	However, to better show the differences between various indices of ACs, we give the received signal expression at $l$-th receive antenna in $r$-th selected AC without Gaussian white noise $\tau_l$, which is expressed as 
	\begin{align} \label{eq:RASM received signal}
		y'_l=\underset{\mathrm{constructive}\,\,part}{\underbrace{\sum_{i=\left( l-1 \right) N_E+1}^{N_E}{h_{2}^{i,l}\Phi _{i,l}h_{1}^{i}}x_k}}+\underset{\mathrm{non}-\mathrm{constructive}\,\,part}{\underbrace{\sum_{q=1,\mathrm{ }q\ne l}^{N_a}{\sum_{i=i'}^{qN_E}{h_{2}^{i,l}\Phi _{i,l}h_{1}^{i}}x_k}}},
	\end{align}
	with  $i'=\left( q-1 \right) N_E+1$, as extended as $\footnote{We assume that each part of REs can perfectly achieve the passive beamforming for specific $l$-th selected antenna as $\varphi _i=\omega _{i,l}+\theta _i$ \cite{ref9}-\cite{ref12}.}$ 
	\begin{align} \label{eq:RASM received signal extend}
		y'_l=\underset{\mathrm{non-constructive}\,\,part}{\underbrace{\sum_{q=1,\mathrm{ }q\ne z}^{N_a}{\sum_{i=i'}^{qN_E}{\beta _{i,l}\alpha _ie^{\jmath\varPsi _{i,l}}x_k}}}}+\underset{\mathrm{constructive}\,\,part}{\underbrace{\sum_{i=\left( l-1 \right) N_E+1}^{lN_E}{\beta _{i,l}\alpha _ix_k}}},
	\end{align}
	where $\Phi _{i,l}$ can be simplified as $\exp \!\:\left( \jmath\varPsi _{i,l} \right) $ representing the adjust phase of the $i$-th RE and for the $l$-th selected receive antenna in selected AC, 
	$\varPsi_{i,l}=\varphi _i-\omega _{i,l}-\theta _i$, 
	$\mathrm{ }l\in \left\{ 1,\mathrm{ }\cdots ,\mathrm{ }N_a \right\} $.
	In the meanwhile, the complete received signal at $l$-th selected antenna is given as $y_l=y'_l+\tau _l$ with $\tau_l$ representing the Gaussian white noise at $l$-th selected receive antenna in selected AC with $\mathcal{C} \mathcal{N} \left( 0,N_0 \right) $ distribution.
	In \equref{eq:RASM received signal} and \eqref{eq:RASM received signal extend}, note that regarding the beam from non-constructive parts and beam narrowness in dense arrays, e.g., $\frac{\lambda}{2}$ spacing, the model incorporates non-constructive reflections, i.e., side lobes and interference, avoiding the assumption of zero interference.
	And the received signal of $u$-th unselected received antenna can be given as
	\begin{align}
		y_u=\sum_{q=1}^{N_a}{\sum_{i=\left( q-1 \right) N_E+1}^{qN_E}{\beta _{i,u}\alpha _ie^{\jmath\left( \varphi _i-\omega _{i,l}-\theta _i \right)}x_k}}+\tau_u,
	\end{align}
	with which the SNR can be obtained as
	\begin{align}
		\gamma _u=\frac{\left\| \sum_{q=1}^{N_a}{\sum_{i=\left( q-1 \right) N_E+1}^{qN_E}{\beta _{i,u}\alpha _ie^{\jmath\left( \varphi _i-\omega _{i,l}-\theta _i \right)}x_k}} \right\| ^2}{N_0}.
	\end{align}
	According to the characteristic of the RIS, when $\varphi _i=\omega _{i,l}+\theta _i$, the SNR at $l$-th selected antenna can be maximized, which is given as
		\begin{align}
			&\left\{ \gamma _l \right\} _{max}=\frac{\left| \sum_{i=\left( l-1 \right) N_E+1}^{lN_E}{\beta _{i,l}\alpha _ix_k} \right|^2}{N_0} \nonumber
			\\
			&+\frac{\left| \sum_{q=1,\mathrm{ }q\ne l}^{N_a}{\sum_{i=\left( q-1 \right) N_E+1}^{qN_E}{\beta _{i,l}\alpha _ie^{\jmath\left( \varphi _i-\omega _{i,l}-\theta _i \right)}x_k}} \right|^2}{N_0}.
		\end{align}
Applying the ML detector, the expression of ML detection in the RASM scheme can be given as follows:
	\begin{align}
		\left\{ \hat{r},\hat{k} \right\} =\mathrm{arg}\min_{r, k} \!\:\left\| \boldsymbol{Y}-\mathbf{G}_{r,k} \right\| _{2}^{2},
		\label{RASM ML}
	\end{align}
	where $\hat{r}$ and $\hat{k}$ represent the estimated $r$-th AC and $k$-th $M$-ary PSK/QAM symbol, 
	and $\mathbf{G}_{r,k} = \mathbf{H}_2\mathbf{\Phi }_r\mathbf{H}_1x_k$.
	Here, $\mathbf{G}_{r,k}$ can be rewritten as $\mathbf{G}_{r,k}=\left\{ G_n \right\} _{n=1}^{N_r}$, 
	where $G_n$ is
	\begin{align}
		G_n=\begin{cases}
			\sum_{i=\left( l-1 \right) N_E+1}^{lN_E}{h_{2}^{i,n}h_{1}^{i}}x_k+G_{n}^{\prime},\,v_{r,n}=1\\
			\\
			\sum_{q=1}^{N_a}{\sum_{i=\left( q-1 \right) N_E+1}^{qN_E}{\beta _{i,u}\alpha _ie^{\jmath\varPsi _{i,l}}x_k}},\,v_{r,n}=0
		\end{cases}
	\end{align}
	with $G_{n}^{\prime}=\sum_{q=1,q\ne l}^{N_a}{\sum_{i=\varsigma}^{qN_E}{h_{2}^{i,n}\Phi _{i,l}h_{1}^{i}}x_k}$.
	
	\vspace{-10pt}
	\subsection{RASSK Transmission}
	In the RASSK scheme, $b_{\mathrm{RASSK}}$ bits are transmitted to the receiver, which are all mapped into the indices of ACs by the receiver. 
	Similar to the above steps, the received signal expression of the RASSK scheme without the constellation symbols can be given as 
	\begin{align} \label{RASSK recevied signal}
		\boldsymbol{Y}=E_s\mathbf{H}_2\mathbf{\Phi }_r\mathbf{H}_1+\mathbf{N},
	\end{align}
	where $E_s$ represents the transmitted energy. The received signal expression without Gaussian white noise at $l$-th receive antenna for the selected AC in the RASSK scheme is
	\begin{align}
		y'_l=&\underset{\mathrm{constructive}\,\,part}{\underbrace{E_s\sum_{i=\left( l-1 \right) N_E+1}^{N_E}{\beta _{i,l}\alpha _i}}}+  \nonumber
		\\
		&\underset{\mathrm{non}-\mathrm{constructive}\,\,part}{\underbrace{E_s\sum_{q=1,\mathrm{ }q\ne l}^{N_a}{\sum_{i=\left( q-1 \right) N_E+1}^{qN_E}{\beta _{i,l}\alpha _ie^{\jmath \varPsi _{i,l}}}}}},
	\end{align}
	which, including Gaussian white noise, can be expressed as $\tau_l$ as $y_l=y'_l+\tau_l$. 
	In the meanwhile, the received signal of $u$-th unselected received antenna in the RASSK scheme is
	\begin{align}
		y_u&=E_s\sum_{q=1}^{N_a}{\sum_{i=\left( q-1 \right) N_E+1}^{qN_E}{h_{2}^{i,u}\Phi _{i,l}h_{1}^{i}}}+\tau_u \nonumber
		\\
		&=E_s\sum_{q=1}^{N_a}{\sum_{i=\left( q-1 \right) N_E+1}^{qN_E}{\beta _{i,u}\alpha _ie^{\jmath\left( \varphi _i-\omega _{i,l}-\theta _i \right)}}}+\tau_u.
	\end{align}
	\par
	Thus, the received SNR at $u$-th unselected antenna is:
	\begin{align}
		\gamma _u=\frac{E_s\left\| \sum_{q=1}^{N_a}{\sum_{i=\left( q-1 \right) N_E+1}^{qN_E}{\beta _{i,u}\alpha _ie^{\jmath\left( \varphi _i-\omega _{i,l}-\theta _i \right)}}} \right\| ^2}{N_0},
	\end{align}
	and then it can be maximized through $\varphi _i=\omega _{i,u}+\theta _i$ at $l$-th selected antenna can be given as
	\begin{align}
		\left\{ \gamma_l \right\}_{\text{max}} &= E_s \frac{
			\left| \sum_{i=(l-1) N_E + 1}^{l N_E} \beta_{i,l} \alpha_i \right|^2
		}{
			N_0 
		} \nonumber \\
		& \quad + E_s \frac{
			\left| \sum_{q=1, q \neq l}^{N_a} \sum_{i=(q-1) N_E + 1}^{q N_E} \beta_{i,l} \alpha_i e^{\jmath\varPsi _{i,l}} \right|^2
		}{
			N_0
		}.
	\end{align}
	The ML detector of the RASSK scheme can be given a
	\begin{align}
		\left\{ \hat{r} \right\} =\mathrm{arg}\min_{r} \!\:\left\| \boldsymbol{Y}-\mathbf{G}_{r} \right\| _{2}^{2}
		\label{RASSK ML}
	\end{align}
	with $\mathbf{G}_{r} = E_s\mathbf{H}_2\mathbf{\Phi }_r\mathbf{H}_1$.
	Meanwhile, $\mathbf{G}_{r}$ can be rewritten as $\mathbf{G}_{r}=\left\{ G_n \right\} _{n=1}^{N_r}$, where $G_n$ can be expressed as
	\begin{align}
		G_n=\begin{cases}
			E_s\left\{ \sum_{i=\left( l-1 \right) N_E+1}^{lN_E}{h_{2}^{i,n}h_{1}^{i}}+G_{n}^{\prime} \right\} , v_{r,n}=1\\
			\\
			E_s\left\{ \sum_{q=1}^{N_a}{\sum_{i=\left( q-1 \right) N_E+1}^{qN_E}{\beta _{i,u}\alpha _ie^{\jmath\varPsi _{i,l}}}} \right\} , v_{r,n}=0\\
		\end{cases}
	\end{align}
	with $G_{n}^{\prime}=E_s\sum_{q=1,q\ne l}^{N_a}{\sum_{i=\varsigma}^{qN_E}{h_{2}^{i,n}\Phi _{i,l}h_{1}^{i}}}$.
	Similar to example of RASM in \figref{system model}, transmitted bit sequence 010 is mapped into 1st and 4th receive antennas, then successfully being detected at receiver.
	
	\subsection{\textcolor{red}{Impact of Quantized RIS Phase Errors}}
	\textcolor{red}{In practical deployments, achieving perfect continuous phase shifts is highly challenging due to hardware impairments and the limited resolution of phase shifters, i.e., finite discrete phase quantization. 
	To account for these practical hardware limitations, a phase error term $\Delta\varphi_{i,m}$ can be introduced. 
	The ideal adjusted phase matrix $\mathbf{\Phi}_r$ is replaced by the practical phase matrix with errors, denoted as $\tilde{\mathbf{\Phi}}_r$, which is given as
	\begin{equation}
		\tilde{\mathbf{\Phi}}_r = \text{diag}\left(e^{j(\varphi_{1,m} + \Delta\varphi_{1,m})}, \dots, e^{j(\varphi_{N,m} + \Delta\varphi_{N,m})}\right),
	\end{equation}
	with $\Delta\varphi_{i,m}$ representing the phase error of the $i$-th reflecting element intended for the $m$-th receive antenna. 
	For a $b$-bit quantized RIS, this phase error is typically modeled as an independent uniform random variable distributed over $[-\frac{\pi}{2^b}, \frac{\pi}{2^b}]$. 
	Consequently, the received signal expressions for the RASM and RASSK schemes, originally given in \equref{RASM recevied signal} and \equref{RASSK recevied signal}, respectively, must be modified to incorporate these hardware impairments:
	\begin{equation}
		\tilde{\boldsymbol{Y}}_{\text{RASM}} = \mathbf{H}_2 \tilde{\mathbf{\Phi}}_r \mathbf{H}_1 x_k + \mathbf{N},
	\end{equation}
	\begin{equation}
		\tilde{\boldsymbol{Y}}_{\text{RASSK}} = E_s \mathbf{H}_2 \tilde{\mathbf{\Phi}}_r \mathbf{H}_1 + \mathbf{N}.
	\end{equation}
	Due to the presence of the phase error $\Delta\varphi_{i,m}$, the perfect passive beamforming condition ,e.g., $\varphi_i = \omega_{i,l} + \theta_i$, required to maximize the SNR cannot be strictly satisfied. }
	\footnote{\textcolor{red}{While this foundational work evaluates the proposed schemes under imperfect CSI and Rician fading, practical deployments face additional impairments like CSI feedback delays, hardware non-idealities, and frequency-selective fading. Incorporating these dynamic factors would render the closed-form analysis mathematically intractable. Therefore, a comprehensive investigation of these practical impairments is left for future research.}}
	
	\section{Generalized Pre-Defined ACs Selection} \label{sec 3}
	Due to the obstruction of the direct link from the RIS to the transmitter and phase adjustment is only optimized for the receiver, obtaining perfect instantaneous CSI from the receiver to the RIS and from the RIS to the transmitter is extremely challenging at the transmitter.
	Meanwhile, estimating instantaneous CSI for each time slot is not high-efficient due to its high computational complexity and additional spectrum resource requirements, which are especially not practical in future wireless networks.
	\par
	\textcolor{green}{Since the total number of mathematically available ACs is $J = 2^{N_r}-1$, but the number of required ACs for bit mapping must be a power of 2, the available ACs are typically more than required. For instance, 3 receive antennas yield 7 non-empty ACs, but only 4 are required to represent 2 bits. Therefore, a generalized pre-defined ACs selection algorithm is essential to eliminate the unnecessary ACs. 
	Before transmission, the transmitter and the receiver pre-agree that exactly $D$ ACs will be utilized to convey the $b_1 = \lfloor \log_2(2^{N_r}-1) \rfloor$ bits, where $D = 2^{b_1}$. These $D$ ACs are chosen from the set of non-empty receive subsets through a specific sequence. In our proposed scheme, we first systematically eliminate the ACs containing the maximum number of antennas, i.e., $N_a = N_r$. Subsequently, we sort the remaining ACs in ascending order based on their $N_a$. Finally, the first $D$ ACs in this sorted list are selected for bit mapping. 
	By employing this method, we not only exclude the ACs with the maximum $N_a$ but also minimize the involvement of ACs with higher $N_a$ in subsequent signal processing stages, which effectively mitigates the impacts of the non-constructive interference. As a concrete example, assuming $N_r=4$, we need to select $D = 2^3 = 8$ ACs from the 15 available ones. By following the aforementioned steps, the single AC with $N_a=4$ and a majority of the ACs with $N_a=3$ are systematically eliminated. The final selected AC set is denoted as $\Lambda(r)$ with $r=1,2,\dots,D$. This designed selection significantly reduces the detection complexity, as detailed in Sec. IV.C.}
	The designed selection can significantly reduce the detection complexity, and a detailed analysis of its detection complexity is provided in sub-section B of \secref{sec 4}.
	The generalized pre-defined ACs selection works at the intelligent unit of the RIS after the information of selected ACs inputting. 
	Then, the intelligent unit controls various parts of REs to amplify the transmitted signal for the specific receive antenna/antennas. 
	\textcolor{red}{By intelligently pruning the transmission states rather than relying solely on post-processing at the receiver, this mechanism intrinsically resolves the complexity bottleneck of dynamic RIS beamforming.}
	
	\vspace{-10pt}
	
	\section{Performance Analysis} \label{sec 4}
	In this section, we discuss the performance analysis of the RASM and RASSK schemes in terms of SE, BER, and detection complexity.
	Firstly, the SE performance of the proposed schemes is presented.
	Notably, to better demonstrate and understand the SE performance, we utilize the bit per channel use (bpcu) of various IM schemes to represent and discuss the SE performance.
	Secondly, we derive and analyze the upper bound of the average BER (ABER) to represent the BER performance of both proposed schemes, assuming Rayleigh channel fading and the use of the ML detector.
	Furthermore, the complexity of ML detector in the proposed schemes and other related schemes is discussed.
	\vspace{-10pt}
	\subsection{SE Performance}
	For each time slot, in the RASM scheme, the SE of $b_1$ bits mapped into ACs can be expressed as
	\begin{align}
		\mathsf{SE_1}=\lfloor \log _2\!\:J \rfloor =\lfloor \log _2\!\:\left( 2^{N_r}-1 \right) \rfloor ,
	\end{align}
	and the SE of $b_2$ bits mapped into constellation symbols can be given as $\mathsf{SE_2}=\log _2\!\:M$. Thus, the total SE of the RASM can be obtained by $SE_1+SE_2$ as
	\begin{align}
		\mathsf{SE_{RASM}}=\lfloor \log _2\!\:\left( 2^{N_r}-1 \right) \rfloor +\log _2\!\:M.
		\label{SE RASM}
	\end{align}
	Meanwhile, the SE of the RASSK scheme can be given as
	\begin{align}
		\mathsf{SE_{RASSK}}=\lfloor \log _2\!\:\left( 2^{N_r}-1 \right) \rfloor .
		\label{SE RASSK}
	\end{align}
	In order to show the superiority in the SE of the RASM and RASSK schemes, we also provide the SE expression of the RSM, RSSK schemes, which are given respectively as follows:
	\begin{align}
		\mathsf{SE_{RSM}}=\log _2\!\:N_r+\log _2\!\:M,
		\label{SE RSM}
	\end{align}
	\begin{align}
		\mathsf{SE_{RSSK}}=\log _2\!\:N_r.
		\label{SE RSSK}
	\end{align}
	Furthermore, the RGSM and RGSSK schemes, as the schemes with higher SE than that of the RSM and RSSK schemes, have a fixed number of selected antennas as $N_s$ in each time slot to convey signals, of which the SE can be given as
	\begin{align}
		\mathsf{SE_{RGSM}}= \bigg\lfloor \log _2\!\:\left( \begin{array}{c}	N_s\\	N_r\\\end{array} \right) \bigg\rfloor +\log _2\!\:M,
		\label{SE RGSM}
	\end{align}
	\begin{align}
		\mathsf{SE_{RGSSK}}=\bigg\lfloor \log _2\!\:\left( \begin{array}{c}	N_s\\	N_r\\\end{array} \right) \bigg\rfloor .
		\label{SE RGSSK}
	\end{align}
	\remark{Comparing \equref{SE RASM} and \equref{SE RASSK}, we can easily find that $\mathsf{SE_{RASM}}>\mathsf{SE_{RASSK}}$ with the same $N_r$, and the SE of both schemes increases with the increase of $N_r$. 
		In order to conduct more impartial SE comparisons, by comparing \equref{SE RASM}, \equref{SE RSM}, \equref{SE RGSM} and \equref{SE RASSK}, \equref{SE RSSK}, \equref{SE RGSSK} respectively, we can draw an insightful conclusion as $\mathsf{SE_{RASM}}\ge \mathsf{SE_{RGSM}}>\mathsf{SE_{RSM}}$ and $\mathsf{SE_{RASSK}}\ge \mathsf{SE_{RGSSK}}>\mathsf{SE_{RSSK}}$ with same $N_r$ and $M$, 
		where the equivalence sign can be applicable when $N_r$ is large enough and $N_s>2$. 
		However, the BER performance of the RGSM and RGSSK schemes has significant deterioration when $N_s>2$, but the RASM and RASSK schemes still exhibit better performance while remaining the same $N_r$.
	\vspace{-10pt}
	\subsection{BER Performance}
	This subsection provides the analytical performance in ABER of both RASM and RASSK schemes, where some insights are also given with $x_k=x_{\hat{k}}=\sqrt{E_s}$ and Rician channel fading with $K=0$ is considered.
	
	\subsubsection{ABER for the RASM Scheme}
	The analytical ABER of the RASM scheme is given.
	According to \equref{RASM ML} of ML detector, the instantaneous pairwise error probability (PEP) can be expressed by giving channel $\mathrm{h}_{1}^{i}$ and $\mathrm{h}_{2}^{i,l}$ as
	\begin{align}
		&\mathrm{P_r}\left( \left\{ r,k \right\} \rightarrow \{\hat{r},\hat{k}\}\mid h_{1}^{i},h_{2}^{i,m} \right) =\mathrm{Pr}\left( \!\:\left\| \boldsymbol{Y}-\mathbf{G}_{r,k} \right\| _{2}^{2} \right) \nonumber
		\\
		&=\mathrm{Pr}\!\:\left( \sum_{n=1}^{N_r}{\left\| y_n-G_n \right\| _{2}^{2}}>\sum_{n=1}^{N_r}{\left\| y_n-\hat{G}_n \right\| _{2}^{2}} \right), 
		\label{RASM PEP1}
	\end{align}
	where $\hat{G}_n$ represents the estimated $G_n$ and $y_n$ represents the received signal at $n$-th receive antenna.
	After some algebraic operations, \equref{RASM PEP1} can be extended and rewritten as
	\begin{align}
		&\mathrm{P_r}\left( \left\{ r,k \right\} \rightarrow \{\hat{r},\hat{k}\}\mid \mathrm{h}_{1}^{i},\mathrm{h}_{2}^{i,m} \right)  =\mathrm{Pr}\!\:\left( \Gamma <0 \right) ,
	\end{align}
	with
	\begin{align}
		\Gamma =\sum_{n=1}^{N_r}{\left\| G_n-\hat{G}_n \right\| _{2}^{2}}+\sum_{n=1}^{N_r}{2\mathfrak{N} \left\{ {\tau _n}^*\left[ G_n-\hat{G}_n \right] \right\}},
	\end{align}
	where ${\tau_l}^*$ represents the complex conjugation of Gaussian white noise and $\Gamma \sim \mathcal{N} \left( \mu _{\Gamma},\sigma _{\Gamma}^{2} \right) $ 
	with $Z=\sum_{n=1}^{N_r}{\left\| G_n-\hat{G}_n \right\| _{2}^{2}}$
	and $\sigma _{\Gamma}^{2}=2N_0\sum_{n=1}^{N_r}{\left\| G_n-\hat{G}_n \right\| _{2}^{2}}$. 
	Thus, by applying the Gaussian Q-function as $Q(\frac{-\mu _{\Gamma}}{\sigma _{\Gamma}^{2}})$ the PEP can be further denoted as follows:
	\begin{align}
		\mathrm{P_r}\left( \left\{ r,k \right\} \rightarrow \{\hat{r},\hat{k}\}\mid \mathrm{h}_{}^{i},\mathrm{h}_{2}^{i,m} \right) 
		=Q\left( \sqrt{\frac{Z}{2N_0}} \right).
	\end{align} 
	Considering the alternative form of Q-function, the unconditional PEP, which is averaged over channel coefficients as $\mathrm{P_r}\left( \left\{ r,k \right\} \rightarrow \{\hat{r},\hat{k}\} \right) =\mathbb{E}_Z\left[ Q\left( \sqrt{\frac{Z}{2N_0}} \right) \right] $, is calculated as
	\begin{align}
		\mathrm{P_r}\left( \left\{ r,k \right\} \rightarrow \{\hat{r},\hat{k}\} \right) &=\int_0^{\infty}{Q\left( \sqrt{\frac{Z}{2N_0}} \right) f_Z\left( Z \right) dZ} \nonumber
		\\&=\frac{1}{\pi}\int_0^{\frac{\pi}{2}}{M_Z\left( \frac{-1}{4\sin ^2\!\:\tau N_0} \right) d\tau},
		\label{RASM MGF}
	\end{align}
	which applies the moment generating function (MGF) of $Z$, given as $M_Z\left( t \right) =\int_0^{\infty}{e^{tZ}dZ}$, with $t=\frac{-1}{4\sin ^2\!\:\tau N_0}$. 
	Actually, $Z$ is the squared Euclidean distance (SED) between the symbols $G_{r,k}$ and $G_{\hat{r},\hat{k}}$. 
	Here, the MGF of \equref{RASM MGF} is derived by the general quadratic form of correlated Gaussian random variables and counts on erroneous or correct detection of the $r$-th and $\hat{r}$-th AC indices. 
	Note that the error in the detection of AC index is able to affect the accuracy of constellation symbol detection. Thus, we separate $Z$ into two categories, which are given as: $Z_1=\left\{ Z \mid r\ne \hat{r} \right\} $ and $Z_2=\left\{ Z \mid r=\hat{r} \right\} $.
	\par
	\noindent 1) First case $Z_1$: $\left\{ Z \mid r\ne \hat{r} \right\} $
	\par
	In this case, the SED can be given as $Z_1=Z_{1}^{1}+Z_{1}^{2}+Z_{1}^{3}$, which are given specifically in \equref{Z11}, \equref{Z12}, and \equref{Z13} with $\varphi _i-\omega _{i,l}-\theta _i=\varPsi_{i,l} $, as shown at the top of next page.
	\begin{figure*}[!] 
		\centering 
		\vspace*{0pt} 
		\begin{align}
			Z_{1}^{1}=\sum\nolimits_{l\in V_r}^{}{\left| \left( \sum_{i=\left( l-1 \right) N_E+1}^{lN_E}{\beta _{i,l}\alpha _i}+\sum_{q=1,\mathrm{ }q\ne l}^{N_a}{\sum_{i=\left( q-1 \right) N_E+1}^{qN_E}{\beta _{i,l}\alpha _ie^{\jmath\varPsi _{i,l}}}} \right) x_k-\sum_{q=1}^{N_a}{\sum_{i=\left( q-1 \right) N_E+1}^{qN_E}{\beta _{i,l}\alpha _ie^{\jmath\varPsi _{i,l}}x_{\hat{k}}}} \right|}^2,
			\label{Z11}
		\end{align}
		
		\begin{align}
			Z_{1}^{2}=\sum\nolimits_{{\hat{l}}\in V_{\hat{r}}}^{}{\left| \sum_{q=1}^{N_a}{\sum_{i=\left( q-1 \right) N_E+1}^{qN_E}{\beta _{i,\hat{l}}\alpha _ie^{\jmath\varPsi _{i,\hat{l}}}x_k}}-\left( \sum_{i=\left( l-1 \right) N_E+1}^{lN_E}{\beta _{i,\hat{l}}\alpha _i}+\sum_{q=1,\mathrm{ }q\ne l}^{N_a}{\sum_{i=\left( q-1 \right) N_E+1}^{qN_E}{\beta _{i,\hat{l}}\alpha _ie^{\jmath\varPsi _{i,\hat{l}}}}} \right) x_{\hat{k}} \right|}^2,
			\label{Z12}
		\end{align}
		
		\begin{align}
			Z_{1}^{3}=\sum_{n=1,n\ne l,n\ne \hat{l}}^{N_r}{\left| \sum_{q=1}^{N_a}{\sum_{i=\left( q-1 \right) N_E+1}^{qN_E}{\left( \beta _{i,n}\alpha _ie^{\jmath\left( \varphi _i-\omega _{i,l}-\theta _i \right)}x_k-\beta _{i,n}\alpha _ie^{\jmath\left( \varphi _i-\omega _{i,l}-\theta _i \right)}x_{\hat{k}} \right)}} \right|^2}.
			\label{Z13}
		\end{align}
		\hrulefill 
	\end{figure*}
	
	Here, $Z_{1}^{1}$, $Z_{1}^{2}$, and $Z_{1}^{3}$ stand for the $m=r$, $m=\hat{r}$, and $m\ne r,\mathrm{ }\hat{r}$ respectively. 
	Then, we can denote $Z_{1}^{1}$ and $Z_{1}^{2}$ as
	\begin{align}
		Z_{1}^{1}=\left| q_1 \right|^2=\left( q_1 \right) _{\mathfrak{N}}^{2}+\left( q_1 \right) _{\mathfrak{T}}^{2},
	\end{align}
	\begin{align}
		Z_{1}^{2}=\left| q_2 \right|^2=\left( q_2 \right) _{\mathfrak{N}}^{2}+\left( q_2 \right) _{\mathfrak{T}}^{2},
	\end{align}
	where $q_1$ and $q_2$ follow complex Gaussian distribution by applying central limit theorem (CLT). 
	$\footnote{Note that $\alpha $ and $\beta $ are the magnitudes of standard complex Gaussian random variables, they follow the Rayleigh distribution with mean value $\frac{\sqrt{\pi}}{2}$ variance $\frac{4-\pi}{4}$, respectively. 
		Moreover, since $\omega _i$, $\varphi _i$, and $\theta _i$ are the phase of standard complex Gaussian random variable and follow the uniform distribution at the range in $(-\pi ,\pi )$. }$
	\footnote{\textcolor{green}{It should be noted that the derived ABER expressions rely on the CLT approximation, which is asymptotically accurate. In practical RIS-assisted systems, the reliability of this approximation depends on the number of reflecting elements $L_i$ assigned to each receive antenna. Through numerical validation, we observe that the approximation remains highly precise when $L_i \ge 8$. As $L_i$ or the total RIS size $N$ increases, the aggregate channel distribution further converges to the complex Gaussian model, making the analytical results more precise.}}
	The mean vector $\boldsymbol{\mu }_1$ and covariance matrix $\boldsymbol{\sigma }_{1}^{2}$ of $\boldsymbol{z}_1=\left[ \left( q_1 \right) _{\mathfrak{R}},\left( q_1 \right) _{\mathfrak{T}},\left( q_2 \right) _{\mathfrak{R}},\left( q_2 \right) _{\mathfrak{T}}\mathrm{ } \right] $, which are respectively given:
	
	\begin{align}
		\boldsymbol{\mu }_1=\frac{N\pi}{4}\left[ \left( x_k \right) _{\mathfrak{N}},\left( x_k \right) _{\mathfrak{T}},-\left( x_k \right) _{\mathfrak{N}},-\left( x_k \right) _{\mathfrak{T}} \right] ^\mathsf{T},
		\label{mean vector}
	\end{align}
	\begin{align}
		\boldsymbol{\sigma }_{1}^{2}=
		\begin{bmatrix}
			\vspace{0.5ex}
			\sigma _{1}^{2}&		\sigma _{1,2}^{2}&	\sigma _{1,3}^{2}&	\sigma _{1,4}^{2}\\
			\vspace{0.5ex}
			\sigma _{1,2}^{2}&	\sigma _{2}^{2}&	\sigma _{2,3}^{2}&	\sigma _{2,4}^{2}\\
			\vspace{0.5ex}
			\sigma _{1,3}^{2}&	\sigma _{2,3}^{2}&		\sigma _{3}^{2}&	\sigma _{3,4}^{2}\\
			\vspace{0.5ex}
			\sigma _{1,4}^{2}&		\sigma _{2,4}^{2}&	\sigma _{3,4}^{2}&	\sigma _{4}^{2}\\
		\end{bmatrix} ,
		\label{variance matrix}
	\end{align}
	where 
	\begin{align}
		\sigma _{1}^{2}=N_a\left[ \left( N_a-\frac{\pi ^2}{16} \right) N_E\left( x_k \right) _{\mathfrak{R}}^{2}+\frac{N\left| x_{\hat{k}} \right|^2\mathrm{ }}{2} \right] ,\nonumber
		\\
		\sigma _{2}^{2}=N_a\left[ \left( N_a-\frac{\pi ^2}{16} \right) N_E\left( x_k \right) _{\mathfrak{T}}^{2}+\frac{N\left| x_{\hat{k}} \right|^2\mathrm{ }}{2} \right] , \nonumber
		\\
		\sigma _{3}^{2}=N_a\left[ \left( N_a-\frac{\pi ^2}{16} \right) N_E\left( x_{\hat{k}} \right) _{\mathfrak{R}}^{2}+\frac{N\left| x_k \right|^2\mathrm{ }}{2} \right] , \nonumber
		\\
		\sigma _{4}^{2}=N_a\left[ \left( N_a-\frac{\pi ^2}{16} \right) N_E\left( x_{\hat{k}} \right) _{\mathfrak{T}}^{2}+\frac{N\left| x_k \right|^2\mathrm{ }}{2} \right] .
	\end{align}
	By applying the property of covariance between two random variables $X$ and $Y$ as
	\begin{align}
		2\mathrm{Cov}\left( X,Y \right) =\mathbb{D}\left[ X+Y \right] -\mathbb{D}\left[ X \right] -\mathbb{D}\left[ Y \right] ,
	\end{align}
	we can easily obtain the $\sigma _{1,2}^{2}$, $\sigma _{3,4}^{2}$, $\sigma _{1,4}^{2}$, $\sigma _{2,3}^{2}$, and $\sigma _{2,4}^{2}$.
	The MGF of the generalized non-central chi-square distribution is given as follows \cite{ref11}-\cite{ref12}:
	\begin{align}
		M_X\left( t \mid \boldsymbol{\mu },\boldsymbol{\sigma }^2 \right)& =\left[ \det \left( \mathbf{E}-2t\boldsymbol{\sigma }^2 \right) \right] ^{-\frac{1}{2}}\times \nonumber
		\\
		&\exp \left\{ -\frac{1}{2}\boldsymbol{\mu }^\mathsf{T}\left[ \mathbf{E}-\left( \mathbf{E}-2t\boldsymbol{\sigma }^2 \right) ^{-1} \right] {\boldsymbol{\sigma }}^{-1}\boldsymbol{\mu } \right\} ,
		\label{closed MGF}
	\end{align}
	where $X=\sum_{f=1}^f{X_{f}^{2}}$, is the unit matrix, $\boldsymbol{\mu }$ and $\boldsymbol{\sigma }^2$ represent the mean vector and covariance matrix of $\left[ X_1,X_2,\cdots ,X_f \right] ^\mathsf{T}$. Then, substituting \equref{mean vector} and \equref{variance matrix} into \equref{closed MGF}, we can yield the MGF of $Z_{1}^{1}+Z_{1}^{2}$, given as $M_{Z_{1}^{1}+Z_{1}^{2}}\left( t \mid \boldsymbol{\mu }_1,\boldsymbol{\sigma }_{1}^{2} \right) $.
	Nevertheless, the MGF of $Z_{1}^{3}$ is still needed to be derived. In \equref{Z13}, we define that
	\begin{align}
		\Gamma =&\sum_{q=1}^{N_a}{\sum_{i=\left( q-1 \right) N_E+1}^{qN_E}{\beta _{i,n}\alpha _ie^{\jmath\left( \varphi _i-\omega _{i,l}-\theta _i \right)}x_k}}  \nonumber
		\\
		&-\sum_{q=1}^{N_a}{\sum_{i=\left( q-1 \right) N_E+1}^{qN_E}{\beta _{i,n}\alpha _ie^{\jmath\left( \varphi _i-\omega _{i,l}-\theta _i \right)}x_{\hat{k}}}},
	\end{align}
	which has variance $\sigma _{\Gamma}^{2}=\frac{N\left( \left| x_k \right|^2+\left| x_{\hat{k}} \right|^2 \right)}{2}$ and zero mean value. 
	For \equref{Z13}, $Z_{1}^{3}$ follows the generalized central chi-square distribution with $2\left( D-2 \right) N_a$ degree of freedom and has MGF:
	\begin{align}
		M_{Z_{1}^{3}}\left( t \right) =\left[ 1-tN\left( \left| x_k \right|^2+\left| x_{\hat{k}} \right|^2 \right) \right] ^{\frac{-N_l}{2}},
	\end{align}
	where $N_l$ represents the number of same antennas in the selected AC and estimated AC.
	Finally, the MGF of $Z_1$ can be obtained as
	\begin{align}
		M_{Z_1}\left( t \right) =M_{Z_{1}^{1}+Z_{1}^{2}}\left( t \mid \boldsymbol{\mu }_1,\boldsymbol{\sigma }_{1}^{2} \right) M_{Z_{1}^{3}}\left( t \right) .
		\label{Z13 MGF}
	\end{align}
	And the unconditional PEP of case 1 can be given by substituting \equref{Z13 MGF} into \equref{RASM MGF} as
	\begin{align}
		&\mathrm{P_r}_{Z_1}\left( \left\{ r,k \right\} \rightarrow \{\hat{r},\hat{k}\} \right) \nonumber
		\\ &=\frac{1}{\pi}\int_0^{\frac{\pi}{2}}{M_{Z_{1}^{1}+Z_{1}^{2}}\left( t\mid \boldsymbol{\mu }_1,\boldsymbol{\sigma }_{1}^{2} \right) M_{Z_{1}^{3}}\left( t \right) d\tau}.
	\end{align}
	\par
	\noindent 2) Second case $Z_2$: $\left\{ Z \mid r=\hat{r} \right\} $:
	\par
	In this case, we also can divide $Z_2$ into $Z_{2}^{1}+Z_{2}^{2}$ with
	\begin{align}
		Z_{2}^{1}=&\left| x_k-x_{\hat{k}} \right|^2\sum_{l\in V}{\left| \sum_{q=1,q\ne l}^{N_a}{\sum_{i=\left( q-1 \right) N_E+1}^{qN_E}{\beta _{i,l}}}\alpha _ie^{\jmath \Psi _{i,l}} \right|^2} \nonumber
		\\
		&+\left| x_k-x_{\hat{k}} \right|^2\sum_{l\in V_r}{\left| \sum_{i=\left( l-1 \right) N_E+1}^{N_E}{\beta _{i,l}}\alpha _i \right|^2},
	\end{align}
	\begin{align}
		Z_{2}^{2}=\left| x_k-x_{\hat{k}} \right|^2\sum_{n=1,n\ne l}^{N_r}{\left| \sum_{q=1}^{N_a}{\sum_{i=i'}^{qN_E}{\beta _{i,n}\alpha _ie^{\jmath\left( \varphi _i-\omega _{i,l}-\theta _i \right)}}} \right|^2},
	\end{align}
	where $Z_{2}^{1}$ and $Z_{2}^{2}$ stand for the situations that $m=r$ and $m\ne r$, respectively. 
	Firstly, by approximating $\beta _{i,l}\alpha _i$ as a Gaussian random variable with CLT, the mean value and variance of $Z_{2}^{1}$ can be obtained as $\mu _{Z_{2}^{1}}=\frac{N_aN_E\pi \left| x_k-x_{\hat{k}} \right|}{4}$ and $\sigma _{Z_{2}^{1}}^{2}=\frac{\left| x_k-x_{\hat{k}} \right|^2N_aN_E\left( 32-\pi ^2 \right)}{16}$. 
	Then, by substituting $\mu _{Z_{2}^{1}}$ and $\sigma _{Z_{2}^{1}}^{2}$ into \equref{closed MGF}, the MGF of $Z_{2}^{1}$ can be obtained as
	\begin{align}
		M_{Z_{2}^{1}}\left( t\mid \mu _{Z_{2}^{1}},\sigma _{Z_{2}^{1}}^{2} \right) =\left( 1-2\sigma _{Z_{2}^{1}}^{2}t \right) ^{-\frac{1}{2}}e^{\frac{{t\mu _{Z_{2}^{1}}}^2}{1-2\sigma _{Z_{2}^{1}}^{2}t}}.
		\label{Z12 MGF}
	\end{align}
	Secondly,  $\Gamma _2$ can be approximately considered as a Gaussian random variable applying CLT:
	\begin{align}
		\Gamma _2=\sum_{q=1}^{N_a}{\sum_{i=\left( q-1 \right) N_E+1}^{qN_E}{\beta _{i,n}\alpha _ie^{\jmath\varPsi _{i,l}}x_k}},
	\end{align}
	with zero mean value and variance $\sigma _{\Gamma _2}^{2}=N$. Thus, $Z_{2}^{2}$ follows the generalized non-central chi-square distribution with $(N_r-N_a)$ degree of freedom and has the variance denoted as $\sigma _{Z_{2}^{2}}^{2}=\frac{NN_a\left| x_k-x_{\hat{k}} \right|^2}{2}$. Giving $\sigma _{Z_{2}^{2}}^{2}$ and following \equref{Z13 MGF}, the MGF of $Z_{2}^{2}$ yields as
	\begin{align}
		M_{Z_{2}^{2}}\left( t \right) =\left( 1-tN_aN\left| x_k-x_{\hat{k}} \right|^2 \right) ^{-\frac{N_r-N_a}{2}}.
		\label{Z22 MGF}
	\end{align}
	Thus, the MGF of $Z_2$ can be obtained by multiplying \equref{Z12 MGF} and \equref{Z22 MGF}, which is specifically given as $M_{Z_2}\left( t \right) =M_{Z_{2}^{1}}\left( t \right) M_{Z_{2}^{2}}\left( t \right) $,
	which can be substituted into \equref{RASM MGF} to further obtain the unconditional PEP of case 2
	$\footnote{Following the upper bound of Q-function, the PEP can be expressed as $\mathrm{P_r}\left( \left\{ r,k \right\} \rightarrow \{\hat{r},\hat{k}\} \right) \le \frac{1}{6}M_Z\left( -\frac{1}{N_0} \right) +\frac{1}{12}M_Z\left( -\frac{1}{2N_0} \right) +\frac{1}{4}M_Z\left( -\frac{1}{4N_0} \right) $ \cite{ref22}. The closed-form expression can be obtained in this way, which is not extended in this paper.}$
	as
	\begin{align}
		\mathrm{P_r}_{Z_2}\left( \left\{ r,k \right\} \rightarrow \{\hat{r},\hat{k}\} \right) =\frac{1}{\pi}\int_0^{\frac{\pi}{2}}{M_{Z_2}\left( t \right) d\tau}. 
	\end{align}
	Eventually, the union bound of ABER for the RASM scheme can be derived from the values of unconditional PEP by \equref{RASM PEP1}, which is given as
	\begin{align} \label{clesed-form RASM}
		P_{b}^{RASM}\le \frac{\sum_r{\sum_{\hat{r}}{\sum_k{\sum_{\hat{k}}{\mathrm{P}_{\mathrm{r}}\left( \left\{ r,k \right\} \rightarrow \left\{ \hat{r},\hat{k} \right\} \right) e'}}}}}{MD\log _2\!\:\left( MD \right)},
	\end{align}
	where $e'=e\left( \left\{ r,k \right\} \rightarrow \left\{ \hat{r},\hat{k} \right\} \right)$ represents the number of bits in error for the corresponding pairwise error event.
	Based on this analytical ABER, more insights of the BER performance can be found with the simulation results in \secref{sec 4}.
	
	\subsubsection{ABER for the RASSK Scheme}
	The analytical ABER performance of the RASSK scheme with the ML detector is given in this sub-section. 
	In light of the steps in ABER derivations of the RASM scheme with the ML detector, the conditional PEP of it can be given:
	\begin{align}
		\mathrm{P_r}\left( r\rightarrow \hat{r} \middle| \mathrm{h}_{1}^{i},\mathrm{ h}_{2}^{i,m} \right) \overset{\mathrm{def}}{=} Q\left( \sqrt{\frac{B}{2N_0}} \right) ,
	\end{align}
	where $B=\sum_{n=1}^{N_r}{\left\| G_n-G_{\hat{n}} \right\|}_{2}^{2}$
	Then, by averaging over channel coefficients, the average unconditional PEP can be obtained as
	\begin{align}
		\mathrm{P_r}\left( r\rightarrow \hat{r} \right) =\frac{1}{\pi}\int_0^{\small{\frac{\pi}{2}}}{M_B\left( \frac{-1}{4\sin ^2\!\:\tau N_0} \right) d\tau}.
	\end{align}
	Since the constellation symbols are not transmitted in the RASSK scheme and $x_k=x_{\hat{k}}=\sqrt{E_s}$., case 1 in the ABER analysis of the RASM scheme with the ML detector is also suitable for the RASSK one where $r\ne \hat{r}$. 
	In other words, the MGF given in \equref{closed MGF} of the RASM scheme in case 1 can be also employed in the RASSK scheme with some modifications in $x_k$ and $E_s$, i.e., $M_B\left( \frac{-1}{4\sin ^2\!\:\tau N_0} \right) \approx M_{Z_1}\left( t \right) $. 
	Finally, the upper bound of the RASSK scheme with the ML detector can be obtained similar to
	\begin{align}  \label{clesed-form RASSK}
		P_{b}^{RASSK}\le \frac{\sum_r{\sum_{\hat{r}}{\mathrm{P}_{\mathrm{r}}\left( r\rightarrow \hat{r} \right) e\left( r\rightarrow \hat{r} \right)}}}{D\log _2\!\:\left( D \right)}.
	\end{align}
	\remark{Since the increase of $N$ leads to the increase of $N_E$ and probably $N_a$, we can find that the results of above MGFs decrease with the arise of $N$, which means that the ABER reduces with the $N$ increase.}
	\remark{\textcolor{red}{The theoretical rigor of this framework is established as follows: First, the ABER expressions in \equref{clesed-form RASM} and \equref{clesed-form RASSK} provide strict theoretical upper bounds derived via the MGF approach. Second, the deterministic even division strategy for RIS phase adjustment circumvents the need for algorithmic convergence proofs, while the ML detector ensures detection optimality. Although employing the CLT for mathematical tractability, the tightness of this asymptotic analysis is robustly validated by Monte Carlo simulations in Sec. \Rmnum{6}.}}
	\vspace{-10pt}
	\subsection{Detection Complexity}
	In this section, the complexity of the ML detector in the RASM and RASSK schemes are analyzed.
	We calculate the complexity based on \equref{RASM ML} in the RASM scheme and \equref{RASSK ML} in the RASSK scheme, where the complex multiplications (CM) and complex additions (CA) are all required to count \cite{ref23}-\cite{ref25}. 
	By doing this, the different steps in \equref{RASM ML} to count the operations of CM and CA in the RASM can be given respectively as follows:
	\begin{itemize}
	\item $N_r(N+1)+N$ operations of CM and $N_r(N-1)+N_r$ ones of CA are needed in $\boldsymbol{Y}$.
	\item Similar to step 1, $\mathbf{G}_{r,k}$ has $N_r(N+1)+N$ operations of CM but only $N_r(N-1)$ ones of CA.
	\item The subtraction operation of $\boldsymbol{Y}-\mathbf{G}_{r,k}$ involves $N_r$ ones of CA.
	\item In the Frobenius norm, $N_r$ CMS and $N_r$ CAs are needed for calculate the magnitude for each element, and $N_r-1$ CAs are required in summing the magnitudes.
	\item In summary, for each set $\left\{ r,k \right\} $, total complexity can be given as $2N_r(N+2)+N$ of CMs and $2N_r(N-1)+4N_r-1$ of CAs.
	\item Lastly, the total number of all set $\left\{ r,k \right\} $ is $D\times M$.
	\end{itemize}
	\par
	Consequently, the total ML detector’s operations of CM and CA in the RASM can be obtained as
	\begin{align}
		C_{RASM}=\left[ 2N_r(2N+1+2N_r)+N-1 \right] DM.
		\label{RASM complexity}
	\end{align}
	As following the same steps of the RASM schemes, $\!\:\left\| \boldsymbol{Y}-\mathbf{G}_{r} \right\| _{2}^{2}$ term has the operations of CM and CA in the ML detector applied in the RASSK scheme can be derived as $2N_r(N+2)+N$ and $2N_r(N-1)+4N_r-1$, respectively. 
	Differing from the RASM scheme, the total operations of the ML detector in the RASSK scheme can be given as
	\begin{align}
		C_{RASSK}=\left[ 2N_r(2N+1+2N_r)+N-1 \right] D.
		\label{RASSK complexity}
	\end{align}
	\remark{In cases where the generalized pre-defined ACs selection is not utilized, $D$ in \equref{RASM complexity} and \equref{RASSK complexity} is replaced into $J$ and $J\gg D$, especially when $N_r$ is large. This leads to a substantial increase in detection complexity, thereby highlighting the significant contribution of the designed ACs selection in reducing complexity.} 
	\remark{Comparing the \equref{RASM complexity} and \equref{RASSK complexity}, we can find the detection complexity of the RASSK schemes is lower than that of the RASM scheme with the same $N_r$ and $N$, because the complexity of the detection in constellation symbols is not considered in the RASSK scheme. Also, according to \equref{RASM complexity} and \equref{RASSK complexity}, the complexity of both schemes increases with the rise in $N_r$ and $N$.}
	\remark{In the RSM and RSSK schemes \cite{ref16}, all receive antennas account for channel fading between the transmitter and the RIS and receive signals reflected from the RIS. 
		The CM and CA operations in these schemes are analogous to those in the RASM and RASSK schemes under the same system model configurations, including $N$ and $N_r$, except for SE. 
		Specifically, the CM and CA operations for RASM and RASSK can be expressed similarly using \equref{RASM complexity} and \equref{RASSK complexity}, with $D$ replaced by $N_r$. This indicates that the complexity of RASM and RASSK approaches that of RSM and RSSK for small $N_r$.  
		Nevertheless, the RASM and RASSK schemes achieve higher SE compared to RSM and RSSK with the same $N_r$. 
		Thus, as the trad-off options applied future wireless networks, we can find that the proposed schemes can yield acceptable complexity and better SE compared to the RSM and RSSK schemes with small $N_r$.}
	\vspace{-10pt}
	
	\subsection{\textcolor{red}{Energy Efficiency Analysis}}
	\textcolor{red}{
	In addition to SE and BER, energy efficiency (EE) is a paramount metric for evaluating the sustainability of modern wireless communication systems, particularly in IoT scenarios. 
	To mathematically formulate the EE for the proposed schemes, we first define the total power consumption model, denoted as $P_{total}$. 
	Unlike traditional active MIMO systems that require multiple power-hungry active RF chains at the transmitter, the proposed schemes rely on a single transmit RF chain and a passive RIS array. 
	Therefore, $P_{total}$ can be modeled as:
	\begin{equation}
		P_{total} = \frac{P_t}{\eta} + P_{TX} + N_r P_{RX} + N P_{RE} + P_{RIS\_ctrl},
	\end{equation}
	where $P_t$ is the actual transmit power, and $\eta$ represents the efficiency of the power amplifier and $P_{TX}$ is the hardware static power consumed by the single RF chain at the transmitter. 
	As all $N_r$ receive antennas remain active to harvest both constructive and non-constructive signals in our schemes, $P_{RX}$ denotes the power consumption per active receive RF chain. 
	For the RIS, $P_{RE}$ represents the near-zero power consumed by each of the $N$ passive reflecting elements for phase shifting, and $P_{RIS\_ctrl}$ is the power consumption of the intelligent controller unit.
	Thus, the EE for the RASM and RASSK schemes can be directly derived by substituting their respective SE from \equref{SE RASM} and \equref{SE RASSK} into the EE definition:
	\begin{equation}
		EE_{RASM} = \frac{\lfloor \log_2(2^{N_r}-1) \rfloor + \log_2 M}{\frac{P_t}{\eta} + P_{TX} + N_r P_{RX} + N P_{RE} + P_{RIS\_ctrl}},
	\end{equation}
	\begin{equation}
		EE_{RASSK} = \frac{\lfloor \log_2(2^{N_r}-1) \rfloor}{\frac{P_t}{\eta} + P_{TX} + N_r P_{RX} + N P_{RE} + P_{RIS\_ctrl}}.
	\end{equation} }
	\remark{\textcolor{red}{It is critical to highlight that the power consumption of a passive RIS element is significantly lower than that of an active RF chain ($P_{RE} \ll P_{TX}, P_{RX}$). 
	By leveraging the RIS to execute adaptive spatial modulation at the receiver side, the RASM and RASSK schemes effectively map extra information bits without activating additional transmit RF chains. 
	Therefore, the numerator (SE) is substantially enhanced while the denominator ($P_{total}$) only experiences a marginal increase from the low-power RIS elements. This architectural advantage guarantees that the proposed schemes can achieve superior EE compared to conventional multiple-active-antenna transmission architectures.}}
	
	\begin{figure}
		\centering
		\includegraphics[width=7.8cm,height=4cm]{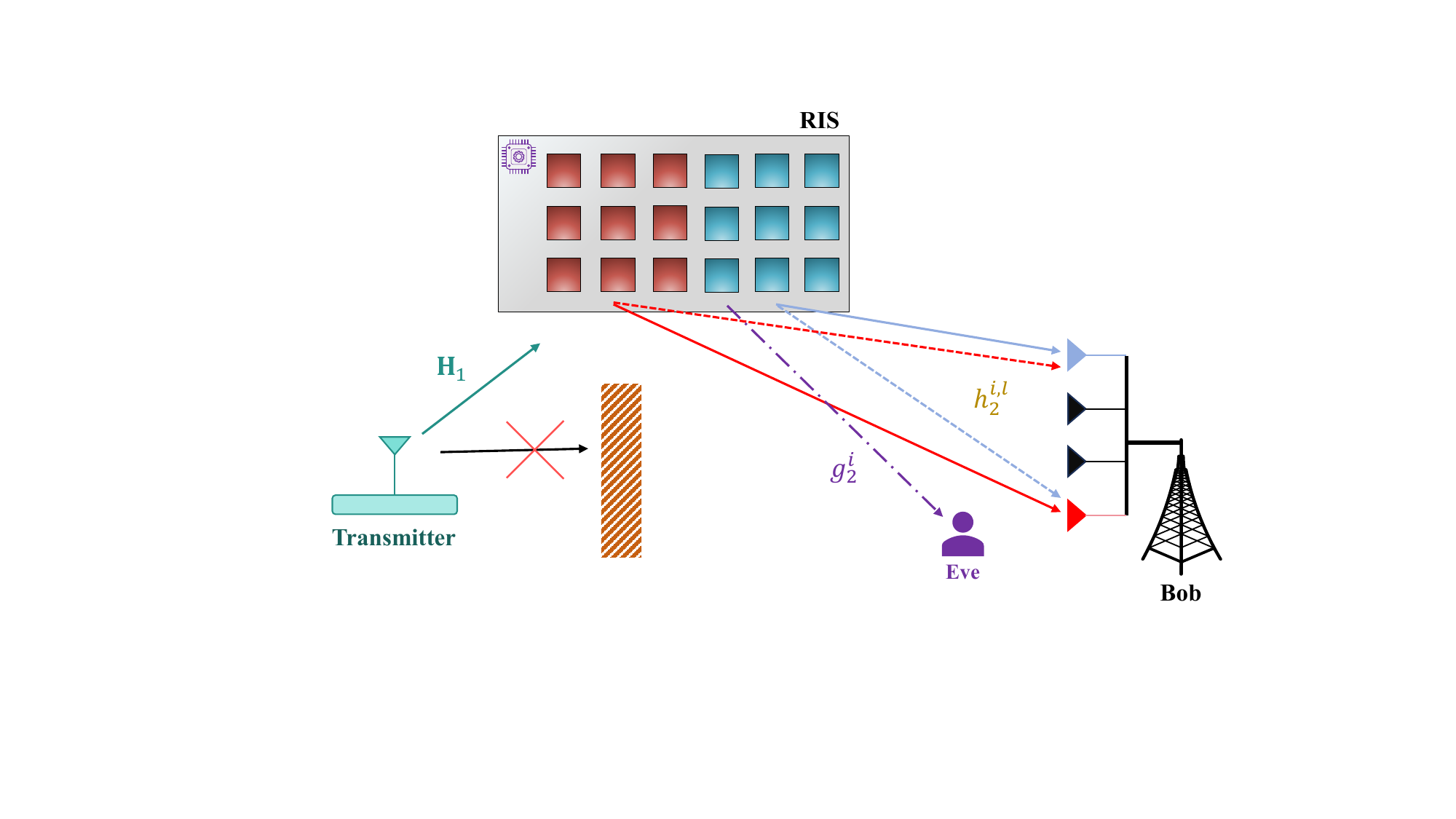}\\
		\caption{System model of the RASM and RASSK schemes eavesdropper.}
		\label{PLS}
	\end{figure}
	
	\section{System Model and Secrecy Rate for PLS} \label{sec 5}
	\subsection{System Model with Eavesdropper}
	Since the adaptive antenna selection works at a specific receiver in both proposed schemes, the eavesdropper cannot completely detect and decode the received signals from the RIS, which further indicates that the proposed schemes offer better secrecy performance in terms of PLS during transmission. 
	A new system model in \figref{PLS} includes an additional eavesdropper, referred to as Eve equipped with one antenna,
	$\footnote{Scenarios involving Eve's physical access to the RIS controller or active jamming, e.g., active eavesdropping or jamming attacks, represent different threat models beyond our scope and are not considered.}$
	which aims at eavesdropping the transmitted signals from the RIS to the receiver as Bob. 
	It is worth noting that this system model in \figref{PLS} is suitable for both the RASM and RASSK schemes, with the only difference that the constellation symbols are transmitted in the RASM scheme instead of the RASSK one. 
	Besides, Eve is assumed to be passive and placed as close to Bob as possible, which means Eve cannot actively attack or interrupt. 
	Here, we also assume that Eve has perfect knowledge of the CSI from the RIS to the receive antenna and the transmitter to the RIS which is given as $g_2\sim \mathcal{C} \mathcal{N} \left( 0,1 \right)$, following Rayleigh distribution \cite{ref26}-\cite{ref28}. 
	Thus, the received signals at Eve can be given as
	\begin{align}
		y_e=\sum_{i=1}^N{g_{2}^{i}\Phi _ih_{1}^{i}x_k}+\tau_e,
	\end{align}
	where $g_{2}^{i}$ and $h_{1}^{i}$ bit sequence $i$-th RE respectively, $\Phi _i$ stands for the adjusted phase from the $i$-th RE, and $n_e$ is the white Gaussian noise with variance $N_0$.  
	\footnote{\textcolor{magenta}{Since $\Phi _i$ is optimized exclusively for Bob's instantaneous channel, it is statistically independent of Eve's channel $g_{2}^{i}$ assuming spatial separation. Consequently, $\Phi _i$ acts as a random phase value for Eve, precluding any coherent beamforming gain in subsequent derivations.}}
	The ML detector at Eve is
	\begin{align}
		\left\{ \hat{k} \right\} =\mathrm{arg}\min_k \!\:\left\| y_e-\Delta _e \right\| _{2}^{2}.
	\end{align}
	where $\Delta _e\overset{\mathrm{def}}{=}\sum_{i=1}^N{g_{2}^{i}\Phi _ih_{1}^{i}x_k}$. 
	In this section, we provide a simple analysis in the SR of the RASM and RASSK schemes.
	\vspace{-10pt}
	\subsection{Secrecy Rate Analysis for the RASM Scheme}
	Referring by \cite{ref29}, the SR of the RASM can be given as
	\begin{align}
		R_{RASM}=\max \!\:\left\{ 0,\mathrm{ }R_B-R_E \right\} ,
		\label{SR form}
	\end{align}
	where $\mathrm{ }R_B$ and $R_E$ represent the data rate of Bob and Eve in bpcu, respectively. 
	Note that each selected AC has the same probability as $\frac{1}{D}$ to be chosen to convey bits sequence in a time slot. 
	Thus, by assuming the data rate for all selected ACs are the same, i.e., $R_{B}^{1}=R_{B}^{2}=\cdots =R_{B}^{D}$ and  $R_{E}^{1}=R_{E}^{2}=\cdots =R_{E}^{D}$, we can easily find that $R_B=R_{B}^{r}$ and $R_E=R_{E}^{r}$, where $R_{B}^{r}$ and $R_{E}^{r}$ represent the Bob’s and Eve’s data rate concerning with $r$-th selected AC. 
	Thus, the data rate of Bob can be further expressed as
	\begin{align}
		R_B=R_{B}^{r}+R_{B}^{k},
		\label{RB total}
	\end{align}
	where $R_{B}^{r}$ and $R_{B}^{k}$ represent the achievable data rate of Bob related to the bits of $r$-th selected AC and $k$-th constellation symbol, respectively. 
	On the contrary, Eve has probability in $\frac{1}{D}$ and $\frac{1}{M}$ to successfully guess and decode the ACs and constellation symbols respectively \cite{ref29}-\cite{ref30}, so the achievable data rate of it can be given as
	$\footnote{Total number of possible mapping patterns for constellation symbols and ACs that the transmitter can adopt are $M$ and $D$, respectively. The probability that Eve adopts the same mapping pattern as the transmitter is $\frac{1}{M}$ for the constellation symbols and $\frac{1}{D}$ for the ACs. This implies that there is only a $\frac{1}{M}$ chance that Eve can correctly guess the exact mapping pattern used by the transmitter for the constellation symbols, and a $\frac{1}{D}$ chance for the ACs.}$
	\begin{align}
		R_E=D^{-1}R_{E}^{r}+M^{-1}R_{E}^{k},
		\label{RE total}
	\end{align}
	with $R_{E}^{r}$ and $R_{E}^{k}$ representing the achievable data rate of Bob related to the bits of $r$-th selected AC and $k$-th constellation symbol when Eve successfully decodes both.
	\par
	According to \cite{ref29}-\cite{ref30}, the expressions for the achievable data rates of Bob and Eve are given in \equref{RB} and \equref{RE} at the top of the next page, respectively. Detailed proofs can be found in \textbf{Appendix A}.
	\begin{figure*}[ht] 
		\centering 
		\vspace*{0pt} 
		\begin{align}
			R_B=\log _2\!\:DM-\frac{1}{DM}\sum_{r=1}^D{\sum_{k=1}^M{\mathbb{E} _{\mathrm{h}_{1}^{i},\mathrm{h}_{2}^{i,n},\tau _n}\left\{ \log _2\!\:\left[ 1+\sum_{\hat{r}=1,\hat{r}\ne r}^D{\sum_{\hat{k}=1,\hat{k}\ne k}^M{\exp \left( -\left\| \boldsymbol{Y}-\mathbf{G}_{r,k} \right\| _{2}^{2}+\mathbf{N} \right)}} \right] \right\}}}.
			\label{RB}
		\end{align}
	\end{figure*}
	\begin{figure*}[ht] 
		\centering 
		\vspace*{0pt} 
		\begin{align}
			&R_E=\frac{1}{M}\left\{ \log _2\!\:M -\frac{1}{M}\sum_{k=1}^M{\mathbb{E} _{n_e,g_{1}^{i},g_{2}^{i}}\left[ \log _2\!\:\frac{1+\exp \left( -\left\| \tau_e \right\| _{2}^{2} \right)}{\sum_{\hat{k}=1}^M{\left[ 1+\exp \left( -\left\| y_e-\sum_{i=1}^N{g_{2}^{i}\Phi _ig_{1}^{i}x_{\hat{k}}} \right\| _{2}^{2} \right) \right]}} \right]} \right\} .
			\label{RE}
		\end{align}
		\hrulefill 
	\end{figure*}
	
	\subsection{Secrecy Rate Analysis for the RASSK Scheme}
	Similar to \equref{SR form}, the SR of the RASSK is given as
	\begin{align}
		R_{RASSK}=\max \!\:\left\{ 0,\mathrm{ }R_B-R_E \right\} ,
	\end{align}
	where $R_B=R_{B}^{r}$ and $R_E=0$ due to the absence of constellation symbols in the RASSK scheme and incapability in detecting the symbols of selected ACs. 
	Furthermore, the mutual information of Bob in the RASSK scheme can be calculated similar to the steps of $\mathfrak{T} _b\left( k;\boldsymbol{Y} \middle| \mathrm{h}_{1}^{i},\mathrm{h}_{2}^{i,n} \right) $ with replacing $x_k$ into $r$-th AC in $\varLambda (r)$ and some algebraic calculations, which can be obtained as $R_{RASSK}=\max \!\:\left\{ 0,\mathrm{ }R_B \right\}$,
	where $R_B=R_{B}^{r}$ and $R_E=0$ due to the absence of constellation symbols in the RASSK scheme and incapability in detecting the symbols of selected ACs. 
	Furthermore, the mutual information of Bob in the RASSK scheme can be calculated similar to the steps of $\mathfrak{T} _b\left( k;\boldsymbol{Y} \middle| \mathrm{h}_{1}^{i},\mathrm{h}_{2}^{i,n} \right) $ with replacing $x_k$ into $r$-th AC in $\varLambda (r)$ and some algebraic calculations.
	Next, taking the expectation of it over $\mathrm{h}_{1}^{i}$ and $\mathrm{h}_{2}^{i,n}$, $R_B$ can be further expressed as
	\begin{align}
		R_B&=\log _2\!\:D- \nonumber
		\\
		&\frac{1}{D}\sum_{k=1}^M{\sum_{\hat{r}=1}^D{\mathbb{E} _{\mathrm{h}_{1}^{i},\mathrm{h}_{2}^{i,n},\tau _n}\left[ \log _2\exp \left( \left\| \mathbf{N} \right\| _{2}^{2}-\left\| \boldsymbol{Y}-\mathbf{G}_r \right\| \right) \right]}}.
		\label{RB2}
	\end{align}
	Substituting \equref{RB2} and $R_E=0$ into $R_{RASSK}=\max \!\:\left\{ 0,\mathrm{ }R_B \right\}$, we can easily find the final SR through Monte Carlo simulations shown in \secref{sec 6}.
	\remark{\textcolor{green}{Our PLS analysis is rooted in strict information-theoretic secrecy, not obfuscation. Following Kerckhoffs's principle \cite{RR_R2_A, RR_R2_B}, we assume Eve possesses full system knowledge, including the alphabet $M$, mapping table $\Lambda$, and RIS strategy. Security fundamentally stems from spatial channel decorrelation: the RIS passive beamforming is exclusively aligned to the legitimate cascaded channel. Consequently, an eavesdropper at a different location experiences an independent, unaligned channel. This physical disparity inherently degrades Eve's mutual information and guarantees a positive secrecy rate, completely independent of her knowledge of the mapping rules.}}
	\remark{\textcolor{magenta}{The RIS phase shifts are strictly optimized for the legitimate cascaded channel. Consequently, even if an eavesdropper intercepts high signal energy, their independent cascaded channel scrambles the delicate energy differences and phase alignments required to decode the spatial activation cues, rendering statistical blind estimation ineffective. This evaluation assumes the ubiquitous condition of independent fading, e.g., $d \gg \lambda/2$. However, if the eavesdropper is located in extreme proximity to the legitimate receiver ($d < \lambda/2$), their channels become highly spatially correlated. Under such specific conditions, significant information leakage is inevitablea fundamental physical limitation governed by electromagnetic wave propagation that is intrinsic to all spatial-domain physical layer security architectures.}}
	
	\begin{figure}
		\centering
		\includegraphics[width=8.5cm,height=8cm]{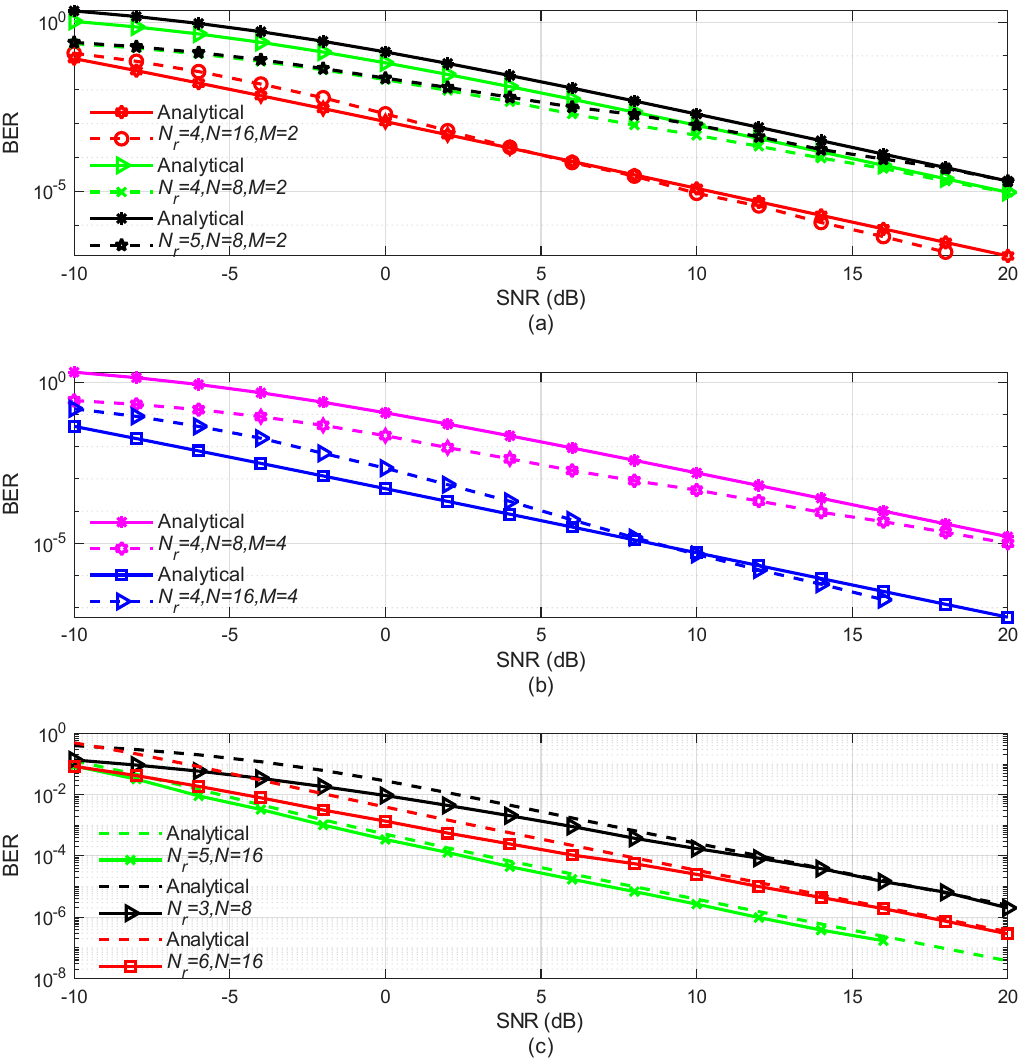}\\
		\caption{Analytical and simulation BER results of the RASM in (a) and (b), and RASSK schemes in (c).}
		\label{Theo}
	\end{figure}

	\vspace{-10pt}
	\section{Simulation Results} \label{sec 6}
	In this section, the analytical and simulation results are presented. 
	To provide a more comprehensive illustration of the performance in the proposed scheme, a uniform assumption of Rayleigh channel fading is adopted, \textcolor{magenta}{except for \figref{Rician},} with the transmission antenna $N_t=1$ to be considered in all cases of the RASM and RASSK schemes. 
	All of these results are conducted by employing the Monte Carlo simulations, encompassing a substantial number of channel realizations, precisely $3\times 10^6$. 
	Similar to \cite{ref10}-\cite{ref14}, $\bar{\gamma}=\frac{E_s}{N_0}$ is considered as the average SNR at receiver.
	\par
	The analytical and simulation results of BER in the RASM scheme are presented in \figref{Theo}, where the results of the cases with $N_r=$ 4 and 5, $N=$8 and 16, $M=$2 and 4 are given, respectively. 
	As depicted in \figref{Theo}, BER decreases monotonically with an increase in $N$. This trend underscores the significant impact of the RIS in enhancing BER performance.
	For the case with $N_r=4$, $N=16$, and $M=2$, we can find that the BER of the RASM scheme is lower than $10^{-5}$ when SNR is equal to 10 dB and bpcu is equal to 4, which indicates the proposed scheme can achieve the desired BER performance. \footnote{\textcolor{magenta}{Since the analytical curves represent theoretical upper bounds derived via the Union Bound technique, these bounds naturally become loose and may mathematically exceed 0.5 in the low-SNR regime due to the summation of overlapping pairwise error probabilities.}}
	Particularly as the SNR increases. the simulation results closely approach to the theoretical ones, which is mainly because of the applications of CLT in theoretical analysis.
	Additionally, the gap between the theoretical and simulation results decreases as $N$ increases. 
	This is also because the CLT utilization, causing the theoretical BER to converge with the simulations as $N$ grows.
	We can observe that the BER performance worsens as $N_r$ increases, which is because the increase in $N_r$ leads to the increase in number of ACs, thus leading to increase error possibility in detection.
	For the same reason, by comparing the results of $M=2$ and $M=4$ between (a) and (b) in \figref{Theo}, the slight discrepancy and learn that the BER performance deteriorates with the increase of $M$.
	\figref{Theo} (c) illustrates the analytical and simulation BER performance of the RASSK scheme, where $N_r=$3, 5, 6 and $N=$8, 16, respectively. 
	Similar to (a) and (b), we learn from (c) that BER performance of the RASSK scheme gets better with the increase of SNR and the analytical BER is approached to the simulation one especially in the higher SNR range. 
	Additionally, the results in (c) also point out that the RASSK has satisfactory BER performance even under the low SNR situation.

	\begin{figure}
		\centering
		\includegraphics[width=8.5cm,height=5.5cm]{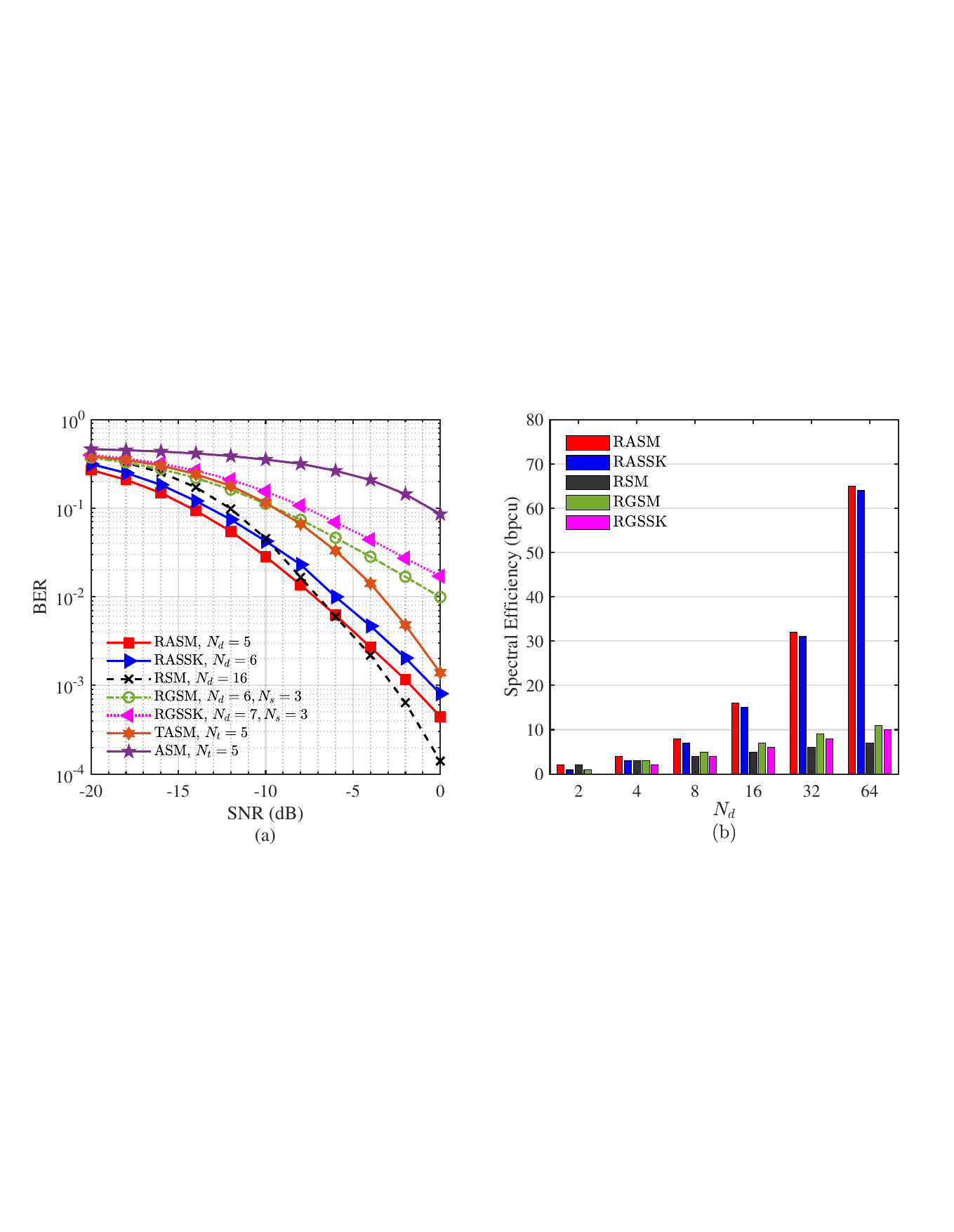}\\
		\caption{(a) BER performance and (b) spectral efficiency represented by bpcu of various IM schemes with different $N_d$, $N_r=16$, $M=2$, and $N=8$.}
		\label{SE}
	\end{figure}

	\par
	Since all of receive antennas are required to receive signals in the received IM schemes, the BER performance gets better with the increase of $N_r$. 
	We only need $N_d$ receive antennas to implement the RASSK-based in antenna selection for the same SE condition during comparisons.
	$\footnote{$N_d$ closely related to SE, which can be easily obtained from \equref{SE RASM}-\equref{SE RGSSK} by replacing $N_r$ into $N_d$. $N_d\ne N_r$ is only given in \figref{SE}. For convenience, we assume $N_d=N_r$ in other figures.}$
	Note that $N_r=16$ is assumed in all received IM schemes of \figref{SE}. 
	For fair comparison, bpcu of all schemes in \figref{SE} (a) is 5, and $N=8$ is given in all schemes of \figref{SE}. Also, $M=2$ is for the RASM, RSM, and RGSM schemes. 
	Thus, the RASM scheme has $N_d=5$, the RASSK scheme has $N_d=6$, the RSM scheme has $N_d=16$, the RGSM scheme has $N_d=6$ and $N_s=3$, and the RGSSK scheme has $N_d=3$ and $N_s=3$, where $N_s$ represents the number of selected antennas in the RGSM and RGSSK schemes \cite{ref18}-\cite{ref19}. 
	Plus, by comparing the BER performance of the TASM and ASM scheme with generalized pre-defined ACs selection, the BER performance of the RASM scheme significantly outperforms that of the TASM and ASM scheme. 
	The reason lies in the fact that within the TASM scheme, the AC indices undergo dual-stage channel fading, contrasting with the single stage in the RASM scheme, while additionally lacking the support provided by the RIS in the ASM scheme. 
	According to \figref{SE} (a), we find that the BER performance of the RASM and RASSK schemes is better than that of the RGSM and RGSSK schemes. 
	Furthermore, the BER performance of the RASM and RASSK schemes is worse than that of the RSM scheme when SNR is larger than around $-2$ dB in the RASM scheme and around -10 dB in the RASSK scheme. However, the BER performance of the RASM and RASSK schemes is better than that of the RSM when SNR is lower than about -2 dB, where the BER performance of the proposed schemes are apparently better with large $N$ and increased $M$.
	In comparison between the RASM and RASSK schemes, we can find that the BER performance of the RASM scheme is better than that of the RASSK, evidencing that the constellation-symbol modulation outperforms the received ACs-selection modulation in the receive IM scheme.
	On the other hand, with the increase of $N_d$, the bpcu of the RASM and RASSK schemes have a significant improvement and is tremendously higher than other schemes especially with the high $N_d$, which further indicates the proposed scheme can provide higher SE in wireless transmission. 
	Since the constellation symbol is not transmitted in the RASSK scheme, the RASM scheme always has 1 bpcu higher than the RASSK scheme when $M=2$, as shown in \figref{SE} (b).  
	\par
	Giving the same condition on SE, the comparison of the various received RIS-assisted IM schemes with the imperfect CSI estimation is shown in \figref{ICSI}, where the schemes include the RASM, RASSK, and RSM schemes. 
	Here, the imperfect CSI is assumed in $h_{2}^{i,m}$ with $\delta  ^2$ in the RTSM and RTSSK schemes representing the variance of estimation error \cite{ref31}-\cite{ref32} and the spectral efficiency is $R=$4 bpcu.
	Detail explanations of the imperfect CSI estimation can be found in Appendix B.
	Results in \figref{ICSI} show that the robustness of the RASSK scheme is better than that of the RASM and RSM schemes, despite the fact that the $\delta  ^2$ in the RASSK results is larger than others. Also, the BER performance of RASM scheme outperforms the one of the RSM scheme.
	\textcolor{red}{Furthermore, when facing hardware-induced phase quantization errors ($b=3$), the RASSK scheme exhibits exceptional resilience, with its BER continuously dropping below $10^{-3}$ at low SNR regions. 
	In contrast, the RASM scheme experiences a noticeable performance floor. 
	This performance gap physically reveals that conveying information purely through received power at selected antennas (as in RASSK) is inherently more robust to phase mismatches than detecting conventional $M$-ary constellation symbols (as in RASM).
	According to these comprehensive comparisons, it can be firmly concluded that the proposed RASSK scheme holds immense potential for applications in complex and imperfect real-world wireless environments due to its superior robustness against both CSI estimation inaccuracies and hardware phase quantization errors. }
	\par
	\figref{Rician} shows BER of RASM scheme with $N_r=8$,$N_d=4$,$N=32$, and $M=2$, RASSK scheme with $N_r=8$ and $N_d=5$, RSM scheme with $N_r=8$, $N=32$, and $M=2$, of which the $K=$5, 10, and 15, respectively.
	In \figref{Rician}, we can find that the BER performance of all schemes gets worse with the increase of $K$.
	This is because stronger LoS component makes the channel responses at different antennas more similar.
	When receive antennas are closely spaced, their channel characteristics become very similar, thus reducing spatial diversity.
	Since IM schemes, especially the SM and SSK related schemes, rely on distinct channel characteristics for correct detection of antenna indices, this reduced diversity leads to degrade BER performance.
	This phenomenon is particularly evident in the results of RASSK scheme shown in \figref{Rician}, which further corroborates the aforementioned argument.
	However, the BER performance of RASM scheme still outperforms the one of RSM scheme even with different $K$.
	\textcolor{magenta}{
	As shown in \figref{Rician}, RASM consistently outperforms RSM across varying $K$. 
	This stems from RASM's adaptive antenna activation, which generates richer spatial signatures and allows a lower constellation order, thereby increasing the minimum Euclidean distance. 
	Also, the RIS phases are optimized based on the instantaneous cascaded CSI, which includes both LoS and Rayleigh components, constructively harvesting Rayleigh energy. 
	Optimizing solely for the LoS path would leave the unaligned Rayleigh component as severe interference. }

	\begin{figure}
		\centering
		\includegraphics[width=8.5cm,height=5.7cm]{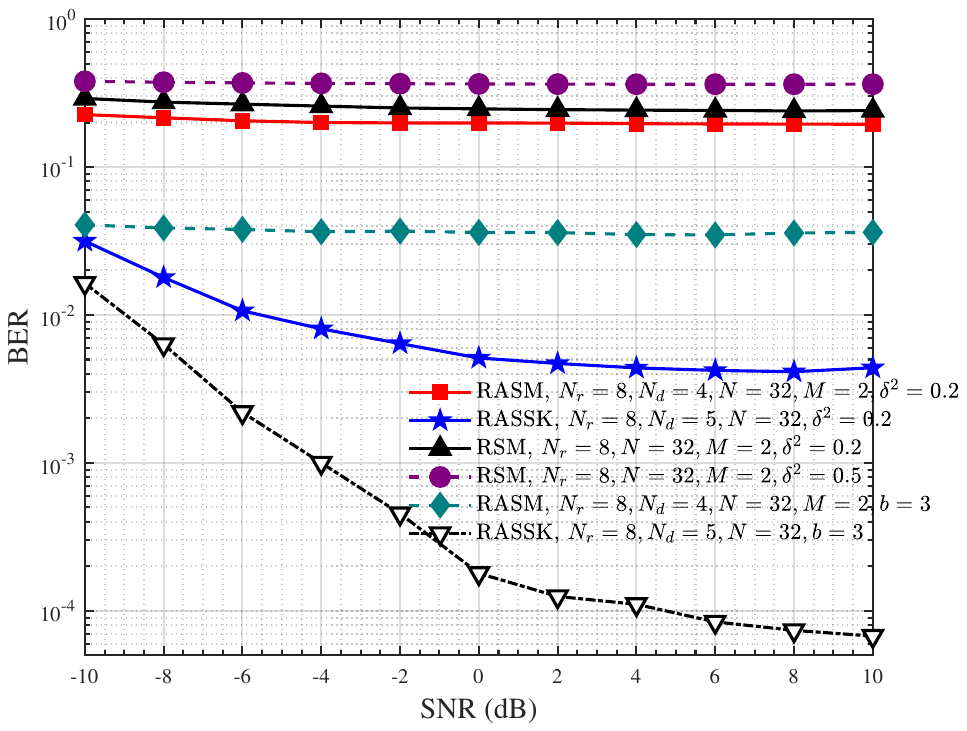}\\
		\caption{\textcolor{red}{BER performance of the RASM, RASSK, and RASM schemes with imperfect CSI estimation and quantized RIS phase errors.}}
		\label{ICSI}
	\end{figure}
	
	\begin{figure}
		\centering
		\includegraphics[width=8cm,height=5.7cm]{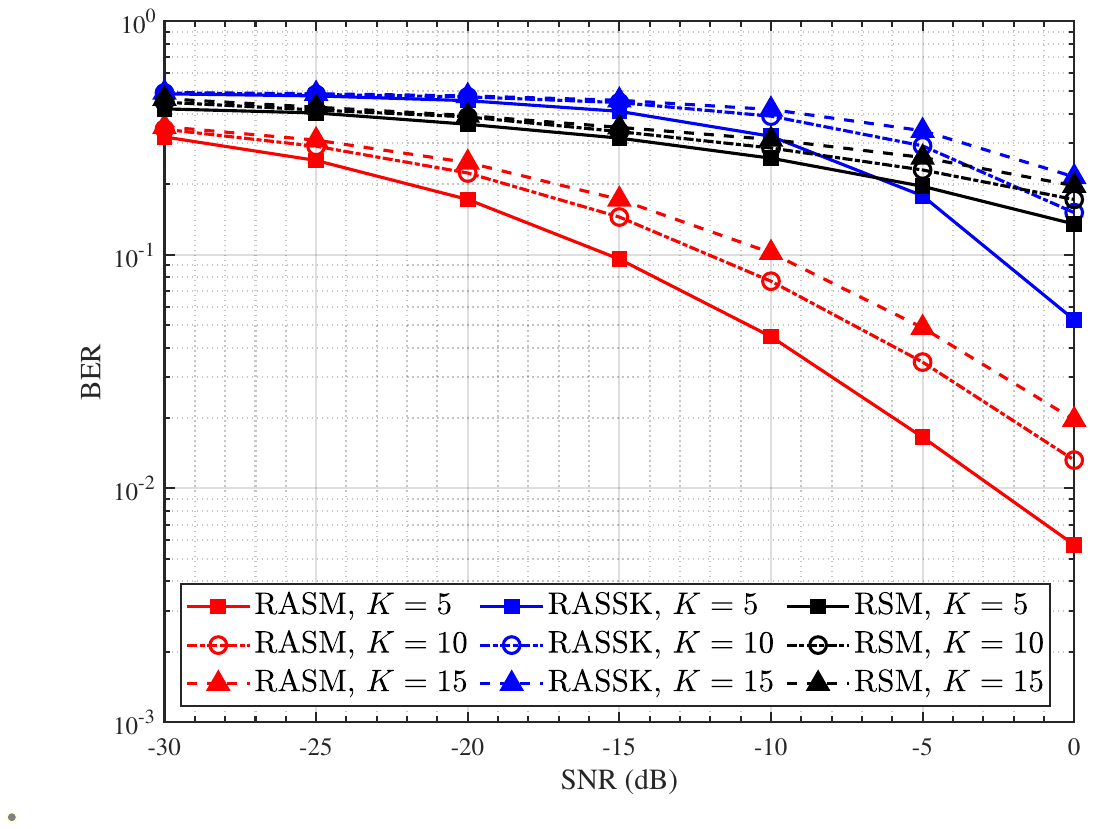}\\
		\caption{BER performance of the RASM, RASSK, and RASM schemes with Rician channel fading in $\mathbf{H}_2$.}
		\label{Rician}
	\end{figure}

	\par
	\figref{SR1} shows the SR of the RASM and RASSK schemes. 
	First, \figref{SR1} shows that the SR of both schemes gets better with the increase of the SNR. 
	Nevertheless, we can also observe that the increase rate stabilizes across all results within the high SNR regime. 
	With the $N$ increasing, the SR of the RASM scheme with $N_r=3$, $N=8$, and $M=2$ is slightly worse than that with $N=16$, which shows the increase of $N$ can improve the PLS performance due to the enhancement in SNR at receiver. 
	In addition, by comparing the results of the RASM scheme with different bpcu, i.e., $N_r=$3, 4 and $M=$2, 4, the SR increases while the bpcu increasing, thus the higher SE provides better PLS performance in the RASM scheme. 
	Specifically, the SR of the RASM scheme with $N_r=3$, $N=16$, and $M=2$ is worse than that of the one with $N_r=4$, $N=16$, and $M=4$, which not only can be learned from the theoretical analysis in \secref{sec 4} where the $R_E$ decreases with the increase in $M$ due to the $\frac{1}{M}$ term in \equref{I5}, but also demonstrates the higher-order modulation is more capable of enhancing the data rate with the assistance of the RIS than the ACs indices. 
	For the RASSK scheme, since it only transmits bits mapped from the ACs selection, the entire signal cannot be successfully detected by Eve.
	Thus, the SR of the RASSK is higher than that of the RASM scheme, which can be obtained by comparing the results of the RASSK scheme and the RASM scheme of the \figref{SR1} (a) and (b), respectively. 
	As also given in \figref{SR1}, it is clear that the SR of the RASSK scheme also increases with the rise of SNR and $N_r$. 
	In contrast to the RASM scheme, the increase rate of SR in the RASSK scheme starts to level off when SNR$<$0 dB, which also can indicate the superiority in PLS of the RASSK scheme. 
	\par
	Following model of MRC in \cite{ref33}, the SR of the RIS-assisted SIMO scheme with MRC (SIMO-MRC), the TSM scheme with MRC (TSM-MRC), the TSM-MRC with AN \footnote{This AN is designed with the noise variance $N_0$ as the same to the Gaussian white noise. Detail AN model can be found in \cite{ref_rr_AN1}.}, and the TSSK scheme with MRC (TSSK-MRC) are given to compare with the one in both proposed schemes as shown in \figref{SR1}.$\footnote{Since there is currently no existing literature discussing and investigating the PLS performance of the TASM scheme, where the adaptive antenna selection works at transmitter and the RIS is employed as a passive relay, we thus select the TSM and TSSK schemes as benchmarks to compare the SR with both proposed schemes in this paper.}$
	To make a fair comparison, as the same SE in 3 bits/s/Hz, the RASM has $N_r=$4, $N=$8, $M=$2, and the RASSK has $N_r=$4, $N=$8, and the SIMO-MRC has $N_t=$1 $N_r=$3, $N=$8, $M=$32, and the TSM-MRC scheme has $N_t=$3 $N_r=$3, $N=$8, $M=$2, and the TSSK-MRC scheme has $N_t=$8 $N_r=$3, $N=$8.
	In \figref{SR1}, the SR of both proposed schemes is higher than that of the others from -30 dB to 30 dB, demonstrating that the received IM schemes achieve better PLS performance compared to the traditional SIMO-MRC, TSM-MRC, and TSSK-MRC schemes, even the PLS wireless technique, e.g., AN in the TSM-MRC.
	Below around -5 dB, the SR of the SIMO-MRC scheme is slightly higher than that of TSM-MRC and TSSK-MRC, indicating that the PLS performance based on transmit antenna indices is weaker than that based on constellation symbols. 
	The SR results of TSM-MRC and TSSK-MRC are closed because Bob and Eve share the same transmitter-to-RIS channel, and the CSI is assumed to be known to Eve, allowing her to fully detect and decode the transmitted signals, which is a limitation of transmitted IM schemes. 
	Notably, the SR of the SIMO-MRC, TSM-MRC, and TSSK-MRC schemes approaches but does not reach zero, as Bob's channel capacity remains higher than Eve's due to having multiple antennas, while Eve only has one.
	\remark{\textcolor{magenta}{Achieving symbol-level RIS reconfiguration demands strict synchronization, as timing failures can cause severe phase misalignment and substantial BER degradation, which is a non-trivial issue and out of our scope, thus leaving for future works. 
		Regarding hardware complexity, both the proposed RASM and conventional fixed-antenna schemes, e.g., RSM, RGSM, and RGSSK, fundamentally require all $N_r$ receive antennas to be connected to active RF chains for spatial detection. 
		Therefore, our adaptive mechanism introduces no additional RF hardware cost at the receiver. 
		This receiver-centric complexity is a deliberate design trade-off, enabling a minimalist, single-RF-chain architecture for power-constrained IoT transmitters.}}
	
	\begin{figure}
		\centering
		\includegraphics[width=8.5cm,height=5.5cm]{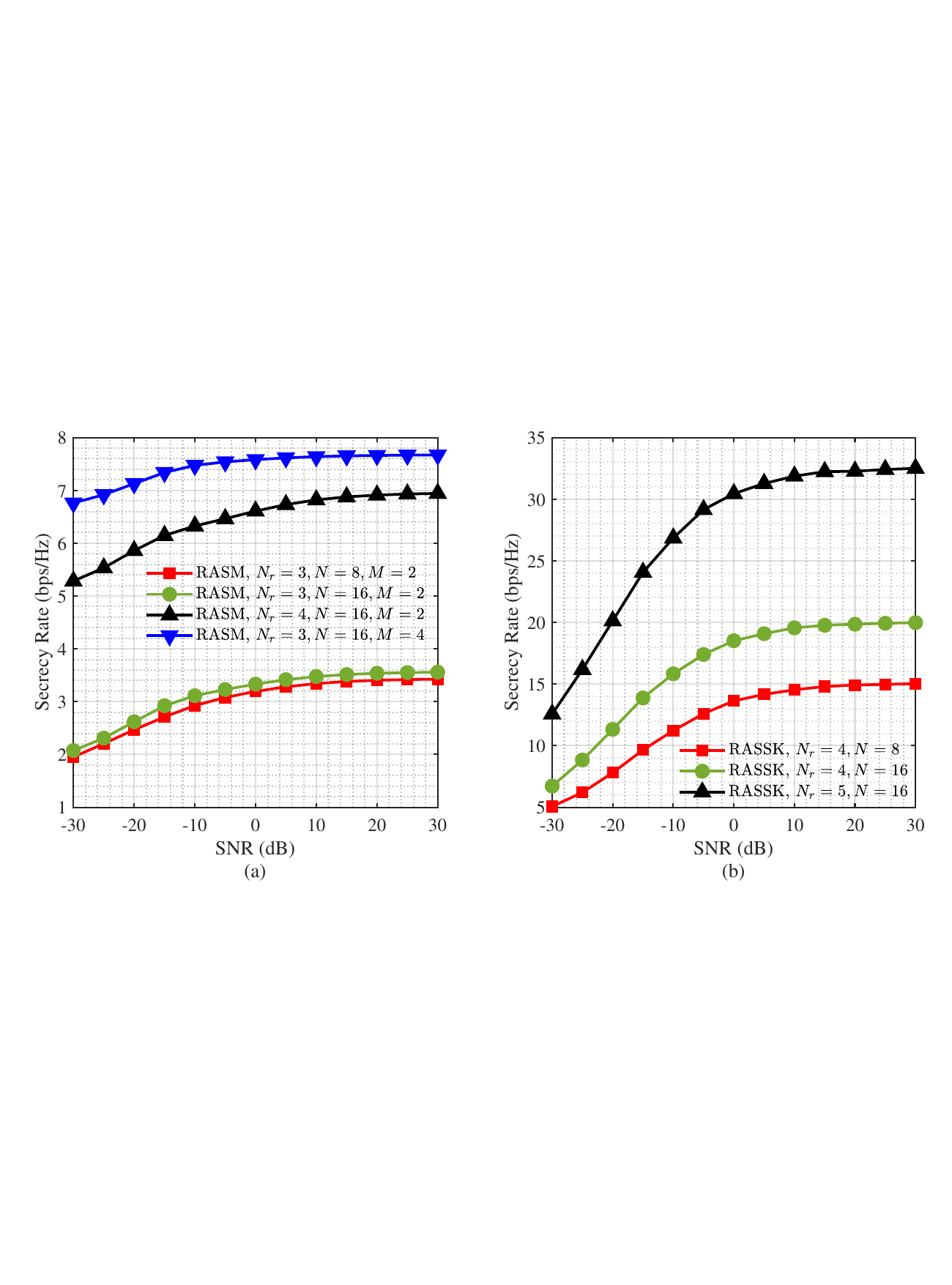}\\
		\caption{(a) SR of the RASM scheme and (b) the RASSK scheme, with $N_r=$3, 4, and 5, $N=$8 and 16, $M=$2 and 4.}
		\label{SR1}
	\end{figure}
	
	\begin{figure}
		\centering
		\includegraphics[width=8cm,height=5.7cm]{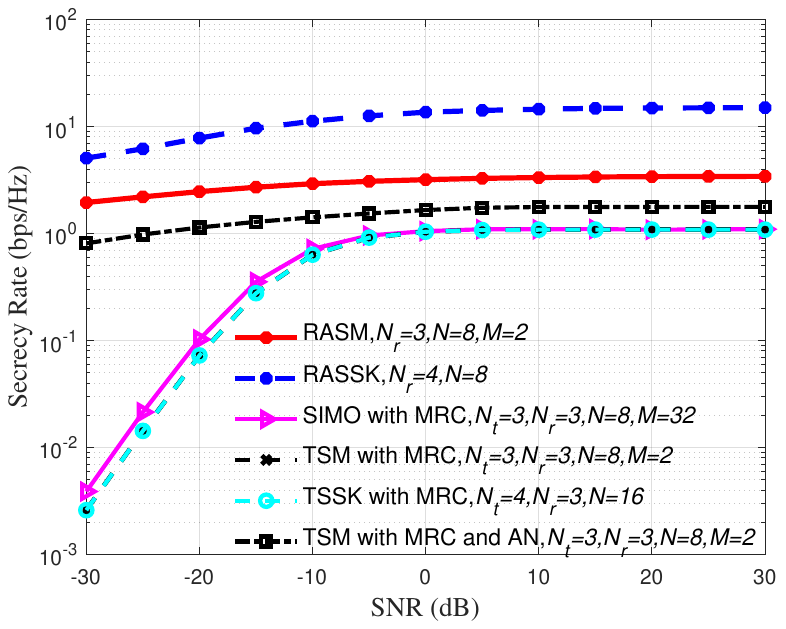}\\
		\caption{SR of the RASM and RASSK schemes compared to the RIS-assisted SIMO, TSM, TSSK schemes empowered by the MRC with $N_r=$3 and 4, $N_t=$1, 4, and 8, $M=$2 and 32, $N=$8, and AN.}
		\label{SR2}
	\end{figure}
	
	\vspace{-10pt}
	
	\section{Conclusion} \label{sec 7}
	This paper presented two novel RIS-assisted IM schemes, RASM and RASSK, aimed at enhancing SE and ensuring PLS in wireless transmissions. 
	\textcolor{magenta}{When the number of receive antennas is 8 or more, the proposed scheme achieves approximately twice the SE of the traditional RSM scheme, while only providing an SNR loss of around 0.5 dB at a target BER lower than $10^{-3}$. 
	Moreover, our physical layer security evaluations confirm that the spatial channel decorrelation inherent to our design strictly guarantees a positive secrecy rate, thoroughly validating the system's robustness against unauthorized eavesdropping.}
	\textcolor{red}{To further approach practical applications, more sophisticated eavesdropper models, as well as the evaluation of the proposed schemes in dynamic multi-user environments with high mobility, will be considered in our future works.}
	Overall, these schemes showed strong potential for future wireless applications.

	\vspace{-10pt}
	\begin{appendices}
		\section{}
		
		The mutual information of Bob and Eve are derived. 
		The achievable data rate for Bob and Eve can be obtained as a discrete-input continuous-output memoryless channel (DCMC) capacity \cite{ref29}-\cite{ref30}. For the RASM scheme, mutual information of Bob can be obtained as:
		\begin{align}
			\mathfrak{T} _b\left( k,r;\boldsymbol{Y} \middle| \mathbf{H}_1,\mathbf{H}_2 \right) =&\mathfrak{T} _b\left( k;\boldsymbol{Y} \middle| \mathbf{H}_1,\mathbf{H}_2 \right) +   \nonumber
			\\
			&\mathfrak{T} _b\left( r;\boldsymbol{Y} \middle| \mathbf{H}_1,\mathbf{H}_2,k \right) ,
			\label{I1}
		\end{align}
		where $y_m$ represents the received signal at $m$-th receive antenna, $\mathfrak{T} _b\left( k;\boldsymbol{Y} \middle| \mathbf{H}_1,\mathbf{H}_2 \right) $ represents the mutual information between $\boldsymbol{Y}$ and $x_k$, $\mathfrak{T} _b\left( r;\boldsymbol{Y} \middle| \mathbf{H}_1,\mathbf{H}_2,k \right) $ stands for the mutual information between $\boldsymbol{Y}$ and $r$-th AC after AC selection when $x_k$ is given. 
		$\mathfrak{T} _b\left( k;\boldsymbol{Y} \middle| \mathbf{H}_1,\mathbf{H}_2 \right) $ can be further expressed as:
		\begin{align}
			&\mathfrak{T} _b\left( k;\boldsymbol{Y} \middle| \mathbf{H}_1,\mathbf{H}_2 \right)  \nonumber
			\\
			&=\sum_{k=1}^M{\int{P_r\left( x_k \right) f\left( \boldsymbol{Y} \middle| k \right)}\log _2\!\:\frac{f\left( \boldsymbol{Y} \middle| k \right)}{f\left( \boldsymbol{Y} \right)}d\boldsymbol{Y}},
			\label{I2}
		\end{align}
		where $P_r\left( x_k \right) =\frac{1}{M}$ since each constellation symbol is selected equiprobably, which is the same in each AC equiprobably selected in $\varLambda (r)$ as $P_r\left( r \right) =\frac{1}{D}$. 
		And the PDF of $\boldsymbol{Y}$ and the conditional PDF of $\boldsymbol{y}$ when $k$ is given are $f\left( \boldsymbol{Y} \right) =\sum_{r=1}^D{\sum_{k=1}^M{\frac{e^{-\!\:\left\| \boldsymbol{Y}-\mathbf{G}_{r,k} \right\| _{2}^{2}}}{DM\pi ^{N_a}}}}$  and $f\left( \boldsymbol{Y} \middle| k \right) =\frac{1}{D}\sum_{r=1}^D{\frac{e^{-\!\:\left\| \boldsymbol{Y}-\mathbf{G}_{r,k} \right\| _{2}^{2}}}{D\pi ^{N_a}}},$ respectively.
		By substituting $f\left( \boldsymbol{Y} \right)$ and $f\left( \boldsymbol{Y} \middle| k \right)$ into (65) and after some algebra, the mutual information between $y_l$ and $x_k$ can be rewritten as
		\begin{align}
			&\mathfrak{T} _b\left( k;\boldsymbol{Y} \middle| \mathbf{H}_1,\mathbf{H}_2 \right) =\log _2\!\:M- \nonumber
			\\
			&\frac{1}{DM}\sum_{r=1}^D{\sum_{k=1}^M{\sum_{\hat{r}=1}^D{\sum_{\hat{k}=1}^M{\mathbb{E} _{\tau _n}\left[ \log _2e^{\!\:\left\| \boldsymbol{Y}-\mathbf{G}_{\hat{r},\hat{k}} \right\| _{2}^{2}-\!\:\left\| \boldsymbol{Y}-\mathbf{G}_{r,k} \right\| _{2}^{2}}\!\: \right]}}}}.
			\label{I3}
		\end{align}
		Also, the mutual information between $y_l$ and $r$-th AC in $\varLambda (r)$ when $x_k$ is given can be obtained as:
		\begin{align}
			\mathfrak{T} _b&\left( r;\boldsymbol{Y} \middle| \mathbf{H}_1,\mathbf{H}_2,k \right) \nonumber
			\\
			&=\sum_{r=1}^D{\sum_{k=1}^M{\int{f\left( k,r \right) f\left( \boldsymbol{y} \middle| k,r \right) \log _2\!\:\frac{f\left( \boldsymbol{y} \middle| k,r \right)}{f\left( \boldsymbol{y} \middle| k \right)}dy_l}}},
			\label{I4}
		\end{align}
		where $f\left( k,r \right) $ represents a joint PDF of $x_k$ and $r$-th AC in $\varLambda (r)$ as a product of their individual PDFs which follows the uniform distribution function, and $f\left( \boldsymbol{y} \middle| k,r \right) $ denotes the conditional PDF of $\boldsymbol{y}$ when $k$ and $r$ are given as:
		\begin{align}
			f\left( \boldsymbol{Y} \middle| k,r \right) =\frac{1}{\pi ^{N_a}}\exp \left( -{\left\| \mathbf{N} \right\| _{2}^{2}} \right) .
		\end{align}
		Substituting $f\left( \boldsymbol{Y} \middle| k \right)$ and \equref{I4} into \equref{I2}, the mutual information between $y_l$ and $r$-th AC in $\varLambda (r)$ is:
		\begin{align} 
			&\mathfrak{T} _b\left( r;\boldsymbol{Y} \middle| \mathbf{H}_1,\mathbf{H}_2,k \right) =\log _2\!\:D-  \nonumber
			\\
			&\frac{1}{DM}\sum_{r=1}^D{\sum_{k=1}^M{\mathbb{E} _{\tau _n}\left[ \log _2\!\:\sum_{r=1}^D{e^{-\left\| \boldsymbol{Y}-\mathbf{G}_{r,k} \right\| _{2}^{2}+\mathbf{N}}} \right]}}.\!\:	
			\label{I5}
		\end{align}
		\par
		By taking expectation of \equref{I5} and \equref{I F1} over $\mathrm{h}_{1}^{i}$ and $\mathrm{h}_{2}^{i,n}$, $R_{B}^{r}$ and $R_{B}^{k}$ can be obtained respectively. 
		Finally, according to \equref{RB total}, the data rate of Bob can be written as \equref{RB}.
		Since Eve only can detect the constellation symbols of the transmitted signals, we can consider that $R_{E}^{r}=0$, thus the \equref{RE total} can be further rewritten as $R_E=\frac{1}{M}R_{E}^{k}$. Besides, assuming both Eve and Bob can perfectly detect the constellation symbols, the mutual information of Eve is similar to $\mathfrak{T} _b\left( k;y_l \middle| \mathrm{h}_{1}^{i},\mathrm{h}_{2}^{i,l} \right) $ as given in \equref{I2}, which can be derived as follows:
		\begin{align}
			\mathfrak{T} _e\left( k;y_e \middle| \mathrm{g}_{1}^{i},\mathrm{g}_{2}^{i} \right) =\sum_{k=1}^M{\int{P_r\left( x_k \right) f\left( y_e \middle| k \right)}\log _2\!\:\frac{f\left( y_e \middle| k \right)}{f\left( y_e \right)}dy_e},
		\end{align}
		with $f\left( y_e \middle| k \right) =\frac{1}{\pi}\exp \left( -\left\| \tau_e \right\| _{2}^{2} \right)$ and
		\begin{align} 
			f\left( y_e \right) =\frac{1}{M}\sum_{k=1}^M{\frac{1}{\pi}\exp \left( -\left\| y_e-\sum_{i=1}^N{g_{2}^{i}\Phi _ih_{1}^{i}x_k} \right\| _{2}^{2} \right)}.
			\label{I F1}
		\end{align}
		Then, by taking expectation of \equref{I F1} over $g_{1}^{i}$ and $g_{2}^{i}$ and multiplying $\frac{1}{M}$, we can obtain the achievable data rate of Eve as \equref{RE}.
		
	\vspace{-10pt}
	\section{  }
	\textcolor{red}{
		The CSI estimation at receiver can hardly be perfect especially in complex wireless environment.
		For simplicity, the imperfect CSI estimation is only considered in the channel from the RIS to receive.
		According to this and referring to the imperfect CSI model from \cite{ref31,ref32}, the coefficient of $\check{h}_{2}^{i,m}$ can be expressed as the sum of $\dot{h}_{2}^{i,m}\sim \mathcal{C} \mathcal{N} (0,1-\delta _{e1}^{2})$ and $\ddot{h}_{2}^{i,m}\sim \mathcal{C} \mathcal{N} (0,\mathrm{ }\delta _{e1}^{2})$.
		Giving that $\delta_{e2}^2$ denotes the inaccuracy of estimation in CSI with $\ddot{\delta_{e2}^2}=1-\delta_{e2}^2$.
		Thus, we can respectively obtain the receive signals at $l$-th receive antenna in $r$-th selected AC for the RASM and RASSK schemes with imperfect CSI estimation as
		\begin{align}
			&\ddot{y}_{l}^{RASM}=\underset{\mathrm{constructive}\,\,part}{\underbrace{\sum_{i=\left( l-1 \right) N_E+1}^{zN_E}{\left( \delta _{e2}^{2}\dot{h}_{2}^{i,m}+\ddot{\delta}_{e2}^{2}\ddot{h}_{2}^{i,m} \right) \Phi _{i,l}h_{1}^{i}}x_k}}+ \nonumber
			\\
			&\underset{\mathrm{non}-\mathrm{constructive}\,\,part}{\underbrace{\sum_{q=1,\mathrm{ }q\ne l}^{N_a}{\sum_{i=\left( q-1 \right) N_E+1}^{qN_E}{\left( \delta _{e2}^{2}\dot{h}_{2}^{i,m}+\ddot{\delta}_{e2}^{2}\ddot{h}_{2}^{i,m} \right) \Phi _{i,l}h_{1}^{i}}x_k}}}+\tau _l,
		\end{align}
		\begin{align}
			&\ddot{y}_{l}^{RASSK}=\underset{\mathrm{constructive}\,\,part}{\sqrt{E_s}\underbrace{\sum_{i=\left( l-1 \right) N_E+1}^{zN_E}{\left( \delta _{e2}^{2}\dot{h}_{2}^{i,m}+\ddot{\delta}_{e2}^{2}\ddot{h}_{2}^{i,m} \right) \Phi _{i,l}h_{1}^{i}}}}+   \nonumber
			\\
			&\sqrt{E_s}\underset{\mathrm{non}-\mathrm{constructive}\,\,part}{\underbrace{\sum_{q=1,\mathrm{ }q\ne l}^{N_a}{\sum_{i=\left( q-1 \right) N_E+1}^{qN_E}{\left( \delta _{e2}^{2}\dot{h}_{2}^{i,m}+\ddot{\delta}_{e2}^{2}\ddot{h}_{2}^{i,m} \right) \Phi _{i,l}h_{1}^{i}}}}}+\tau _l.
		\end{align}
		In the meanwhile, the received signals with imperfect CSI estimation of $u$-th unselected received antenna in the RASM and RASSK schemes can be respectively given as:
		\begin{align}
			&\ddot{y}_{u}^{RASM} \nonumber
			\\
			&=\sum_{q=1}^{N_a}{\sum_{i=\left( q-1 \right) N_E+1}^{qN_E}{\left( \delta _{e2}^{2}\dot{h}_{2}^{i,m}+\ddot{\delta}_{e2}^{2}\ddot{h}_{2}^{i,m} \right) \Phi _{i,l}h_{1}^{i}}}x_k+\tau _u,
		\end{align}
		\begin{align}
			&\ddot{y}_{u}^{RASSK} \nonumber
			\\
			&=E_s\sum_{q=1}^{N_a}{\sum_{i=\left( q-1 \right) N_E+1}^{qN_E}{\left( \delta _{e2}^{2}\dot{h}_{2}^{i,m}+\ddot{\delta}_{e2}^{2}\ddot{h}_{2}^{i,m} \right) \Phi _{i,l}h_{1}^{i}}}+\tau _u.
		\end{align}	
	}
\end{appendices}
	
\end{document}